%% file: elsarticle-template-num.tex
\documentclass[preprint,12pt]{elsarticle}

\usepackage{amssymb}
\rmfamily{\fontsize{12pt}{\baselineskip}\selectfont}

\usepackage{graphicx}
\usepackage{amsfonts, amsmath, amssymb, amsbsy, amsthm}
\usepackage{bm}
\usepackage{color}
\usepackage{xcolor}
\usepackage{hyperref}
\usepackage{mathrsfs}
\usepackage{longtable}
\usepackage[top=2.5cm,bottom=2.5cm,left=2.5cm,right=2.5cm]{geometry}
\usepackage{epstopdf}
\usepackage{stmaryrd}
\usepackage[normalem]{ulem}
\usepackage{cancel}
\usepackage{scalerel}
\usepackage{enumitem}
\usepackage{algorithm}
\usepackage{subfigure}
\usepackage{algorithm, algorithmicx}
\usepackage[noend]{algpseudocode}
\usepackage{lineno}
\usepackage{enumitem}
\usepackage{diagbox}
\usepackage{multirow}
\renewcommand{\algorithmiccomment}[1]{\bgroup\hfill//~#1\egroup}

\newtheorem{theorem}{Theorem}[section]
\newtheorem{assumption}{Assumption}[section]
\newtheorem{lemma}{Lemma}[section]
\newtheorem{remark}{Remark}[section]

\newcommand\tbbint{{-\mkern -16mu\int}}

\newcommand\dbbint{{-\mkern -19mu\int}}

\newcommand\bbint{
{\mathchoice{\dbbint}{\tbbint}{\tbbint}{\tbbint}}
}

\input{notation}

\journal{}

\begin{document}

\begin{frontmatter}

\title{A Theoretical Case Study of the Generalisation of Machine-learned Potentials}

\author[ubcaddress]{Yangshuai Wang\corref{mycorrespondingauthor}}
\cortext[mycorrespondingauthor]{~Corresponding author. \textbf{Funding}: YW is supported by Development Postdoctoral Scholarship for Outstanding Doctoral Graduates from Shanghai Jiao Tong University. }
\ead{yswang2021@math.ubc.ca}

\author[shashwataddress]{Shashwat Patel}

\author[ubcaddress]{Christoph Ortner}

\address[ubcaddress]{Department of Mathematics, University of British Columbia, Vancouver V6T1Z2, Canada.}

\address[shashwataddress]{Department of Metallurgical and Materials Engineering, Indian Institute of Technology Madras, Chennai, Tamil Nadu, India}

\begin{abstract}
Machine-learned interatomic potentials (MLIPs) are typically trained on datasets that encompass a restricted subset of possible input structures, which presents a potential challenge for their generalization to a broader range of systems outside the training set. Nevertheless, MLIPs have demonstrated impressive accuracy in predicting forces and energies in simulations involving intricate and complex structures. In this paper we aim to take steps towards rigorously explaining the excellent observed generalisation properties of MLIPs. Specifically, we offer a comprehensive theoretical and numerical investigation of the generalization of MLIPs in the context of dislocation simulations. We quantify precisely how the accuracy of such simulations is directly determined by a few key factors: the size of the training structures, the choice of training observations (e.g., energies, forces, virials), and the level of accuracy achieved in the fitting process. Notably, our study reveals the crucial role of fitting virials in ensuring the consistency of MLIPs for dislocation simulations. Our series of careful numerical experiments encompassing screw, edge, and mixed dislocations, supports existing best practices in the MLIPs literature but also provides new insights into the design of data sets and loss functions.
\end{abstract}



%

\end{frontmatter}


\section{Introduction}
\label{sec:intro}
Machine-learned interatomic potentials (MLIPs)~\cite{2019-ship1, Bart10, behler07, Braams09, Drautz19, Shapeev16} for atomistic materials simulations have gained widespread attention in the past decade and now becoming part of the standard toolbox of computational materials science.
The achievement of MLIPs is to bridge the significant gap in accuracy and capability between ab initio electronic structure models \cite{kohanoff2006electronic, kotliar2006electronic, saad2010numerical} and classical mechanistic models (empirical potentials) \cite{Daw1984a, LJ, Stillinger1985Computer}. 

A key question in machine learning tasks is to understand how well a trained model generalizes to inputs outside of the training data. This is particularly challenging in scientific machine learning where one oftentimes requires generalisation to inputs {\it far from} training data (out of distribution). Several recent works~\cite{de2022error, mishra2022estimates} proposed to study the generalization of physics-informed neural networks (PINN) in solving partial differential equations (PDEs). 

In the present work we study the generalization of MLIPs for atomistic materials simulations. In this case, MLIPs are typically trained on limited data-sets that cover only a very small section of the full space of possible input structures containing at most hundreds of atoms. Predictions during simulations are on much larger domains often containing hundreds of thousand or even millions of atoms. As a paradigm case study, we selected the MLIPs simulation of dislocations in crystalline solids. This setting is sufficiently well understood that it can be studied by rigorous analytical tools, while still being of significant interest for materials modelling. 

The investigation of dislocations has a long history in materials modelling due to the substantial influence that dislocations exert on the mechanical, electronic, and thermal material properties~\cite{cai2003periodic, eshelby1953anisotropic, lardner1971mathematical}. By undertaking a comprehensive study of dislocation generation, motion, and interaction, one can attain a deeper and more comprehensive understanding of the mechanical response and failure modes exhibited by materials. Employing MLIPs in simulating dislocations opens up new opportunities for improved quantitative understanding how they influence material behaviour~\cite{dragoni2018achieving, grigorev2023calculation,  hodapp2020operando}. 

In a recent work~\cite{ortner2022framework}, we propose a generalisation analysis framework for point defect simulations to explain how the choice of training data and the accuracy of the fit to that training data affect the accuracy of predictions. As opposed to the commonly used statistical method to study the uncertainty quantification and error propagation~\cite{liang2011error, roy2011comprehensive}, this is an analytical method reminiscent of the {\it a priori} error analysis of multiscale schemes~\cite{2021-qmmm3, mlco2013}. In this setting, the various approximation errors can be classified and the influence of various approximation parameters on the error made precise. 
While \cite{ortner2022framework} outlines a general framework for such a generalisation analysis, it is limited only to the study of point defects. 
While it is clear that an extension to more complex scenarios is conceptually possible, the intricacies involved in such an extension require a thorough investigation.

Our primary contribution in the present work lies in providing such an extension for dislocation simulations. By leveraging known properties of their associated cores and elastic fields we will provide rigorous qualitative estimates of MLIPs generalisation errors. In contrast to point defects studied in~\cite{ortner2022framework}, the presence of dislocations introduces additional complexities in estimating errors and constructing approximation parameters. The slower decay of the elastic far-field associated with dislocations results in a fundamental distinction: training solely on energies and forces is insufficient to achieve a {\it consistent} MLIP for dislocation simulations; it is crucial to incorporate accurate fits to linear (and nonlinear) elastic response. 
Aside from the intrinsic value of a rigorous theoretical analysis, our study results in interesting practical considerations for optimizing the training of MLIPs, which we will highlight throughout this work and summarize in the Conclusion.

\subsection*{Outline}
We focus on multiple dislocations with periodic boundary conditions, where a rigorous numerical analysis approach is in principle feasible. The atomistic equilibration problem for a single crystalline defect in this context is a well-defined variational problem~\cite{chen19, Ehrlacher16}. In Section~\ref{sec:models} we review the framework and adapt it to the case of multiple dislocations considered in this work along the line of~\cite{2014-dislift}.

In Section \ref{sec:gen}, we demonstrate how the accuracy of basic properties such as defect geometry and formation energy in the simulation depends explicitly on the size of the training structures, on the kind of observations (energies, forces, elastic constants and virials) to which the model has been fitted, and on the training accuracy. For the sake of simplicity of notation, we limit the rigorous analysis on screw dislocations. However, we are confident from related works~\cite{Ehrlacher16} that the topological aspects of edge and mixed dislocations do not change these results but only introduce additional technical complications. 
We will numerically verify these more general cases and leave the analysis to future work. The explicit theoretical convergence rates are summarized in Theorem~\ref{them:geometry} and Table~\ref{table-e-mix}. 

Subsequently, we present a practical implementation of MLIPs based on the insights gained from our generalization analysis. This implementation is tested and validated on several model problems, including screw, edge, and mixed dislocations, as described in Section~\ref{sec:numer}. 

Finally, we will discuss further consequences and limitations of our work  in Section \ref{sec:conclu}. For example, the extension to incommensurate 2D materials or extrapolating on grain boundary structures with distinct coordination environment, there are additional challenges that our analysis does not cover even heuristically and requires additional ideas.

\subsection{Notation}
We use the symbol $\langle\cdot,\cdot\rangle$ to denote an abstract duality
pairing between a Banach space and its dual. The symbol $|\cdot|$ normally
denotes the Euclidean or Frobenius norm, while $\|\cdot\|$ denotes an operator
norm.
For a finite set $A$, we will use $\#A$ to denote the cardinality of $A$.
For the sake of brevity of notation, we will denote $A\backslash\{a\}$ by
$A\backslash a$, and $\{b-a~\vert ~b\in A\}$ by $A-a$.
For $E \in C^2(X)$, the first and second variations are denoted by
$\<\delta E(u), v\>$ and $\<\delta^2 E(u) v, w\>$ for $u,v,w\in X$.
For $j\in\N$, ${\bm{g}}\in (\R^d)^A$, and $V \in C^j\big((\R^d)^A\big)$, we define the notation
\begin{eqnarray*}
	V_{,{\bm \rho}}\big({\bm g}\big) :=
	\frac{\partial^j V\big({\bm g}\big)}
	{\partial {\bm g}_{\rho_1}\cdots\partial{\bm g}_{\rho_j}}
	\qquad{\rm for}\quad{\bm \rho}=(\rho_1, \ldots, \rho_j)\in A^{j}.
\end{eqnarray*}
The symbol $C$ denotes a generic positive constant that may change from one line
of an estimate to the next. When estimating rates of decay or convergence, $C$
will always remain independent of the system size, the configuration of the lattice and of the test functions. The dependence of $C$ will be normally clear from the context or otherwise stated explicitly.
The closed ball with radius $r>0$ and center $x$ is denoted by $B_r(x)$, or $B_r$ if the center is the origin.

\section{Background}
\label{sec:models}
\setcounter{equation}{0}

\subsection{Dislocation far fields}
\label{sec:sub:defects}
Our theoretical results will utilize known properties of equilibrium dislocation configurations, which we now review. These results will {\em not} form part of the computational schemes, but only used to analyze model errors.

We consider a model for straight dislocation lines 
following the setup of~\cite{Ehrlacher16}. Let $\mathsf{B}\mathbb{Z}^3$ denote a 3D Bravais lattice oriented in such a way that the dislocation direction can be chosen parallel to $e_3$ and the Burgers vector can be chosen as $\bv=(\bv_1,0,\bv_3) \in \mathsf{B}\mathbb{Z}^3$. We further assume, without loss of generality, that the displacement fields are independent of the $x_3$-direction and thus only functions of $x_1$ and $x_2$.  We denote the resulting (projected) 2D reference lattice by
\[
\Lambda = \mA \mathbb{Z}^2 := \{ (\ell_1, \ell_2): \ell \in \mathsf{B}\mathbb{Z}^3 \}.
\]
We restrict our analysis to single-species Bravais lattices. While there are no conceptual obstacles to generalising our analysis to multi-lattices, the notational details become more involved. Our numerical exploration in Section~\ref{sec:numer} will also include tests in a multi-lattice setting.

Let $u^{\rm CLE}(\cdot; \bv): \L\rightarrow\R^3$ denote a {\it far-field predictor} for single straight dislocation with Burgers vector $\bv$, solving the associated continuum linearised elasticity (CLE) equation~\cite{anderson2017theory}. The derivation of $u^{\rm CLE}$ is reviewed in the~\ref{sec:apd:u0}.

A general deformed configuration of the infinite lattice $\L$, with single straight dislocation configuration, is a map $y: \L\rightarrow\R^3$, decomposed into
\begin{eqnarray}\label{y-u}
y(\ell) = \ell + u^{\rm CLE}(\ell) + u(\ell) = y^{\rm CLE}(\ell) + u(\ell),
\end{eqnarray}
where the {\it relative displacement} field $u: \L\rightarrow\R^3$ is called the {\it core corrector} and accounts for discreteness and nonlinearity in atomistic models.

Let $\bar{y}^{\L}$ be the equilibrium state (under a suitable atomistic interaction law, to be specified later) of the single straight dislocation configuration of the infinite lattice $\L$ and $\bar{u}^{\L}:=\bar{y}^{\L}-y^{\rm CLE}$. Under mild and general conditions on the lattice and interaction law, it was rigorously shown~\cite{chen19, Ehrlacher16} that the equilibrium core corrector has a generic decay, 
\begin{eqnarray}\label{eq:reg_single_infinite}
    \big|D\bar{u}^{\L}(\ell)\big| \leq C |\ell|^{-2}\log(|\ell|),
\end{eqnarray}
where $D\bar{u}^{\L}(\ell)$ is a a finite difference gradient of $\bar{u}^{\L}$ centered at $\ell\in\L$. The rigorous definition of $Du(\ell)$ is given in~\eqref{eq: nn norm}. 
%
A consequence of the decay \eqref{eq:reg_single_infinite} is that one can define a truncation operator $\Pi_R$ such that $\Pi_R u(\ell) = 0$ for all $\ell \in \L \setminus B_R$ with approximation error 
\begin{equation} \label{eq:truncation operator:main}
    \| D \Pi_R \bar{u}^\L - D \bar{u}^\L \|_{\ell^2} 
    \leq C R^{-1} \log({R}).
\end{equation}
The construction of $\Pi_R$ is subtle~\cite[Section 7.2]{Ehrlacher16} and is therefore reviewed in \eqref{eq:trun_op_def}.

\subsection{Supercell simulations}
\label{sec:sub:supercell}
When simulating dislocations using electronic structure models like density functional theory (DFT), periodic boundary conditions are commonly used. In the case of periodic domains (supercells), it is necessary to consider a periodic array of dislocations with alternating signs. 
We therefore now {\it formally} extend the predictor-corrector framework reviewed in the previous section to multiple dislocations in a supercell (instead of an infinite domain) and will then use this setting in the remainder of the paper to study training and generalisation of MLIPs.
For the sake of simplicity of presentation, we will skip over some technical details but fill these gaps in the~\ref{sec:apd:pre}.

We specify the {\em simulation} domain as a continuous cell $\Omega_N:=\mathsf{P}(-N/2, N/2]^2$, where $\mathsf{P}=(p_1, p_2) \in \R^{2 \times 2}$ is invertible and $p_1, p_2 \in \mathsf{A}\Z^2$. For a sufficiently large $N \in \N$, let 
\[
\L_N:=\L\cap\Omega_N \quad \textrm{and} \quad \L_N^{\per}:= \bigcup_{\alpha \in N\Z^2}(\mathsf{P}\alpha + \L_N),
\]
where $\L_N$ is the supercell and $\L_N^{\per}$ is the resulting periodically repeated infinite defective lattice. We denote the space of periodic displacements by
\[
\Us^{\per}_N := \{u:\L_N^{\per} \rightarrow \R^3~|~u(\ell+\mathsf{P}\alpha) = u(\ell) ~\textrm{for}~\alpha \in N\Z^2\}.
\]

\begin{figure}[!htb]
    \centering
    \includegraphics[height=6cm]{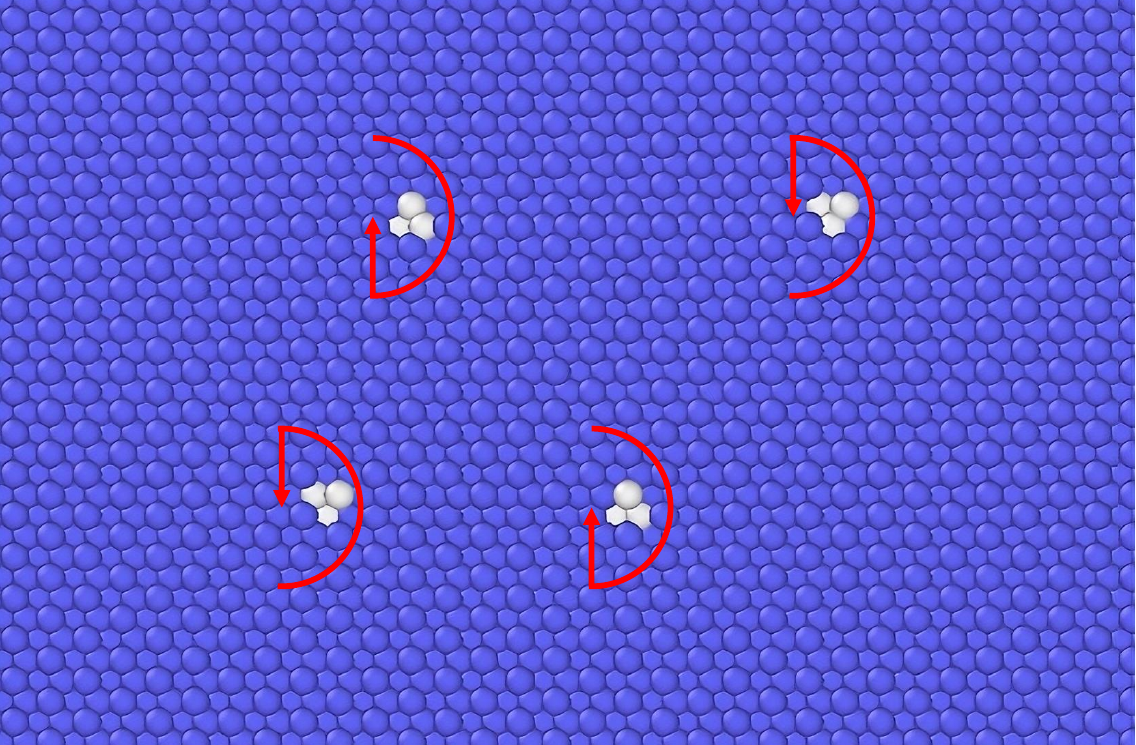}
    \caption{Illustration of a configuration of four screw dislocations in W, colored via Common Neighbor Analysis in Ovito~\cite{stukowski2009visualization}.}
    \label{fig:illuW}
\end{figure}

We define a {\em dislocation configuration} to be a set $\mathcal{D}$ of $n_{\D}$ pairs $(x^{\rm core}_i, \bv_{i})\in \Omega_N \times \mathsf{B}\mathbb{Z}^3$, where $x^{\rm core}_i$ is the core position for the $i$-th dislocation with accompanying Burgers vector $\bv_i$. For compatibility with periodic boundary condition, we require that $n_{\D}$ is even and the net-Burgers vector vanishes, i.e., $\sum_{i} \bv_i={\bm 0}$. An illustration of a quadrupole screw dislocation configuration in W is shown in Figure~\ref{fig:illuW}. We define the minimum separation distance of $\D$ by 
$$
L_{\D} := \min_{i\neq j} |x^{\rm core}_i - x^{\rm core}_j|.
$$ 

Analogous to $u^{\rm CLE}$,  
we denote the CLE predictor for a periodic dislocation configuration $\D$ by
$u^{\rm CLE}_{\rm per}(\cdot; \D): \L_N^{\rm per} \rightarrow\R^3$; for the details of its construction see~\cite{cai2003periodic}. As in \eqref{y-u} we can decompose a periodic configuration into $y(\ell) = \ell + u^{\rm CLE}_{\rm per}(\ell) + u(\ell)$, where $u$ is now the corrector field for multiple cores. We will now use the knowledge about core structure in the infinite-domain decomposition \eqref{y-u} to further refine the construction in the periodic case.

More specifically, let $\E(y)$ be a potential energy functional defined acting on a periodic configuration $y$. An equilibrium defect geometry is obtained by solving
\begin{align}\label{eq:variational-problem-per}
    \bar{y} \in \arg\min \big\{ \E(y), 
    ~y-y^{\rm CLE}_{\rm per} \in \Us_N^{\per} \big\},
\end{align}
where $y^{\rm CLE}_{\rm per}(\ell) := \ell+u^{\rm CLE}_{\rm per}(\ell)$.
%
Motivated by the theory for a single straight dislocation~\eqref{eq:reg_single_infinite}, we now make an intuitive and mild assumption on the existence, stability and regularity of equilibrium configurations for \eqref{eq:variational-problem-per}. 

\begin{assumption}\label{ass:existence}
    Let $\D$ be a dislocation core configuration with sufficiently large core separation distance $L_{\D} \geq L_0$. 
    Then we assume that there exists a strongly stable equilibrium $\bar{y}$ of \eqref{eq:variational-problem-per} satisfying 
\[
\exists~c_0 >0 \quad {\rm s.t.}\quad \<\delta^2\E(\bar{y})v, v\> \geq c_0\|Dv\|^2_{\ell^2(\L_N)},
\]
with a stencil norm $\|\cdot\|_{\ell^2}$ defined by~\eqref{eq: nn norm}. Moreover, we assume that the equilibrium $\bar{y}$ can be decomposed into
\begin{eqnarray}\label{eq:eq_decomp_}
\bar{y}(\ell) = y^{\rm CLE}_{\rm per}(\ell) + \sum_{(x^{\rm core}_i, \bv_i) \in \D} \Pi_{R} \bar{u}^{\L}(\ell-x^{\rm core}_i; \bv_i) + \omega(\ell), \qquad \forall \ell \in \L_N,
\end{eqnarray}
where the defect core truncation operator $\Pi_R$ is defined by \eqref{eq:trun_op_def} with radius $R=L_{\D}/3$, and the reminder term $\omega$ satisfies 
\begin{eqnarray}\label{eq:reminder_estimates}
\|D\omega\|_{\ell^2(\L_N)} \leq C \,\sqrt{n_{\D}} \,L_{\D}^{-1}\log(L_{\D}),
\end{eqnarray}
with a constant $C$ dependent on the interaction law, on the stability constant $c_0$, but independent of the dislocation configuration $\D$ (except, possibly, implicitly through $c_0$), the separation distance $L_\D$, or the domain size $N$. 
\end{assumption}

Assumption~\ref{ass:existence} not only gives the existence of the equilibrium of the multiple dislocations in a periodic domain, but also establishes its structure: The equilibrium can be decomposed into two parts, a truncated defect core centered at each point defect and a remainder term. Our generalisation analysis in the next section heavily relies on this result.
The condition that $L_{\D}$ is sufficiently large entails that defect cores do not overlap too strongly and hence the truncated infinite-lattice core corrector provides a good estimate for the core structure of interacting dislocations. The scaling of $L_{\D}^{-1}\log(L_{\D})$ of the remainder $\omega$ is directly related to the truncation of the core at radius $R \propto L_{\D}$; cf. \eqref{eq:truncation operator:main}. 
The scaling $\sqrt{n_{\D}}$ of the remainder is simply due to accounting for the number of cores being truncated in a 2-norm.

\begin{remark}\label{re:assump}
We believe that Assumption~\ref{ass:existence} can be proven rigorously, possibly requiring some additional assumptions to avoid edge cases about the distribution of dislocation cores. Similar results for point defects are proven in \cite[Theorem 2.1]{ortner2022framework} and for dislocation with different boundary conditions and simplified interaction law in~\cite{2014-dislift}. To give further evidence for Assumption~\ref{ass:existence} we give a sketch of a proof in~\ref{sec:apd:as}. This proof assumes a uniform bound on the possible number of dislocation cores but is otherwise general. Removing that bound would be the main technical hurdle to a fully rigorous proof.

Since the focus of the current work is on a model error analysis for machine learned interatomic potentials, a rigorous proof of Assumption~\ref{ass:existence} goes well outside the scope of the current paper.  Even if Assumption~\ref{ass:existence} were only valid under additional assumptions, this would not change the relevance of our following main results. 
\end{remark}

\subsection{Machine-learned interatomic potentials (MLIPs)}
\label{sec:sub:mlips}
%
%
Small-scale single-defect simulations, requiring only few evaluations (e.g. geometry optimisation) can be routinely carried out using an electronic structure model such as density functional theory. However, due to the significant computational expense associated with electronic structure models, large-scale multi-defect simulations and long-time evolution are normally undertaken using interatomic potentials. While those were mostly empirical models in the past, it is now possible to construct machine-learned interatomic potentials (MLIPs)~\cite{2019-ship1, Bart10, behler07, Shapeev16}, fitted to an electronic structure model, and closely matching its predictions. MLIPs are becoming part of the standard toolbox of computational materials science. Our general analysis is agnostic to the choice of MLIP architecture, hence we give only a brief and generic introduction.

Virtually all modern MLIPs for materials model the total energy $\mathcal{E}^{\rm ML}$ as a sum of site energies, 
\[
\mathcal{E}^{\rm ML}(y; \pmb{c}) = \sum_{i} \mathcal{E}_i^{\rm ML}(y; \pmb{c}),
\]
where $\mathcal{E}^{\rm ML}_i$ describes the local energy contribution from the $i$-th atomic site. The site energy $\mathcal{E}_i^{\rm ML}(y; \pmb{c})$ is parameterized, and optimization of its parameters $\pmb{c}$ is achieved through the minimization of a loss function. 

Given a training set $\mathfrak{R}$ containing atomic configurations $y_R$, together with corresponding observations: total energies $\E(y_R)$, forces $-\nabla \E(y_R)$, and possibly other quantities such as virials, hessians, and so forth. A common choice in materials modelling is a quadratic cost function penalizing errors in energy, forces is 
\begin{eqnarray}\label{eq:common_loss}
\mathcal{L}(\pmb{c}) := \sum_{y_R \in \mathfrak{R}}\Big(W_{\rm  E}\big|\E(y_R) - \E^{\rm ML}(y_R; \pmb{c})\big|^2 + W_{\rm F}\big|\nabla \E(y_R) - \nabla \E^{\rm ML}(y_R; \pmb{c}) \big|^2\Big),
\end{eqnarray}
where $W_{\rm E}, W_{\rm F}$ are weights that may depend on the configurations as well as the observations.

We will loosely think of \eqref{eq:variational-problem-per} as the ``high-fidelity'' model, too expensive to solve in practice. 
Given an MLIP $\mathcal{E}^{\rm ML}$ fitted to $\mathcal{E}$, we can now instead compute the equilibrium geometry with the new potential energy model,  
\begin{align}\label{energy-difference-per-ML}
\bar{y}^{\rm ML} \in& \arg\min \big\{ \E^{\rm ML}(y),~y-y^{\rm CLE}_{\rm per} \in \Us_N^{\per} \big\}.
\end{align}
%
The geometry and energy errors committed in the approximate problem \eqref{energy-difference-per-ML} are, respectively, 
\[
\|D\bar{y} - D\bar{y}^{\rm ML}\|_{\ell^2(\L_N)} \qquad \textrm{and} \qquad \big| \E(\bar{y}) - \E^{\rm ML}(\bar{y}^{\rm ML}) \big|.
\]
The selection of training data, loss functions, and weight parameters leading to \eqref{eq:common_loss} plays a pivotal role in achieving accurate MLIPs (i.e. with the ability to make accurate predictions). The primary focus of this paper is to provide an analytical framework dedicated to elucidating these critical choices within the context of dislocation simulations.

\section{Generalization and Error estimates}
\label{sec:gen}
In this section we develop a theoretical framework to assess the error between the exact and approximate equilibrium geometries $\bar{y}, \bar{y}^{\rm ML}$ and energies $\mathcal{E}(\bar{y}), \mathcal{E}^{\rm ML}(\bar{y}^{\rm ML})$ in terms of the selected training data and in terms of the accuracy of the fit. 
Since MLIPs are fitted to actual {\it ab initio} data, the training domains must be chosen small (order 10s to 100s or atoms). We think of the ability to train on small domains but predict on large and complex structures as a form of generalisation, which we can study rigorously within our framework. 
To that end, we first introduce the training domains, the {\it matching conditions} between the ({\it ab initio}) reference and the MLIPs, and then give a rigorous error estimate for predictions on large simulation domains with (potentially) large and complex dislocation configurations.

\subsection{Training domains and matching conditions}
\label{sec:sub:gen}
In the setting of Assumption~\ref{ass:existence} (dislocation cores cannot get too close) it is intuitive that each training domain should only contain a single dislocation. However, in contrast to our approach for point defects as presented in \cite{ortner2022framework}, a dipole configuration needs to be taken into consideration to enable the use of periodic boundary conditions on the training domain. 

\begin{figure}[!htb]
    \centering
    \includegraphics[height=6cm]{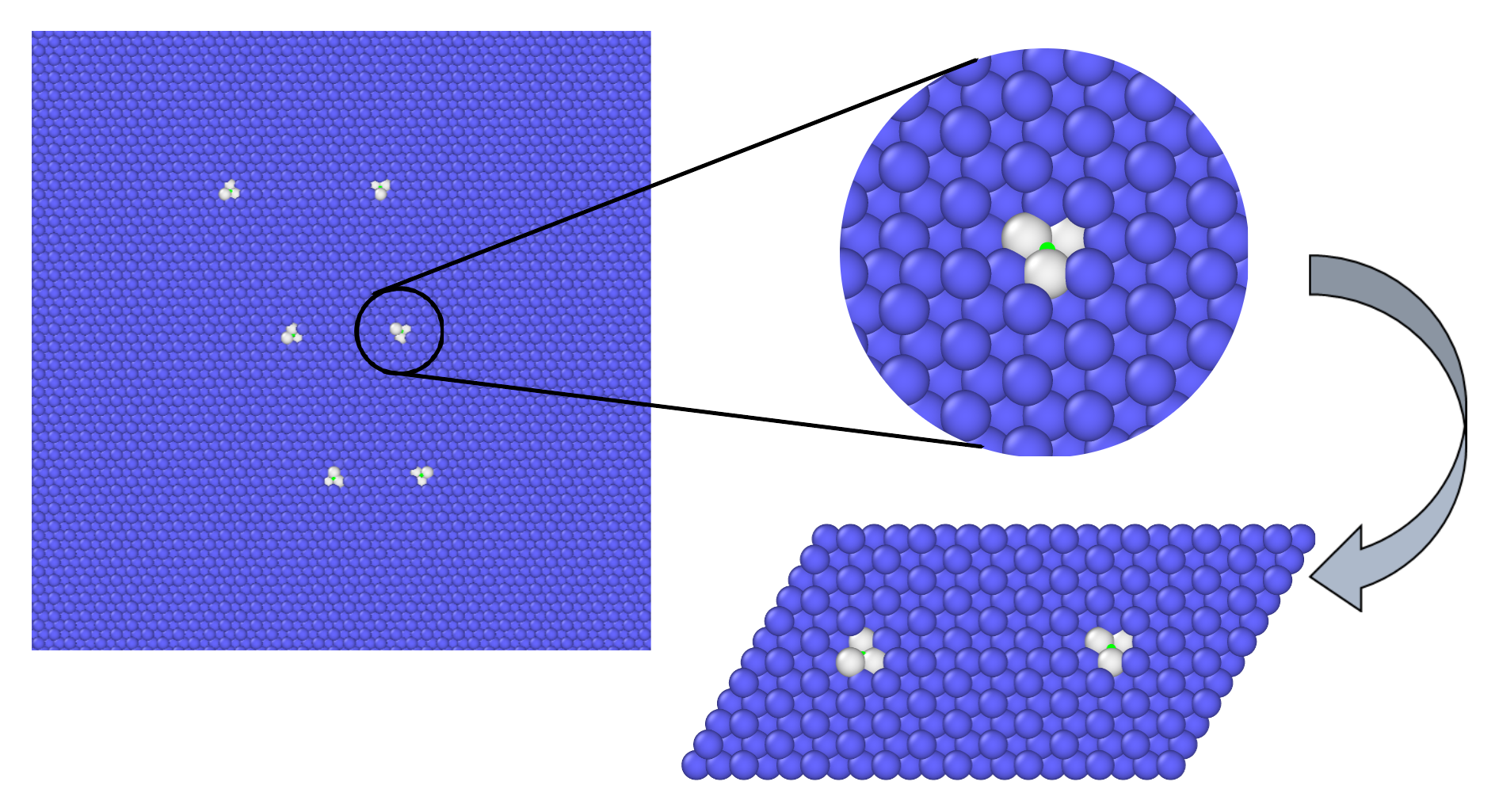}
    \caption{Illustration of simulation domain $\Omega_N$ (left) and training domain $\Omega_L$ (lower right) for screw dislocations in W. }
    \label{fig:illu_W_dipole}
\end{figure}

To be more precise, given $L \leq L_{\D} \ll N$, we call $\Omega_L:=\mathsf{P}(-L, L]^2$ the {\it training domain} while $\Omega_N$ is called the {\it simulation domain}. The defected lattice $\L_L:= \L \cap \Omega_L$ incorporates a dipole configuration to facilitate the application of periodic boundary conditions (see e.g.~\cite{hodapp2020operando}) with distance of order $O(L)$ to prevent interaction. 
See Figure~\ref{fig:illu_W_dipole} for an illustration of screw dislocations in W. Similarly as in the previous section, let $\Us^{\rm per}_{L}$ be the corresponding space of periodic displacements. We equip $\Us_L^{\rm per}$ with the norm $\|v\|_{\Us_L^{\rm per}} := \|Dv\|_{\ell^2(\L_L)}$. Let $\E_L(y)$ and $\E^{\rm ML}_L(y)$ be the energy functionals defined on $\L_{L}$, and the equilibrium of the corresponding variational problem with $\E_L(y)$ is denoted by $\bar{y}_L$. 
To train an MLIP capable of accurately simulating dislocations we ensure that the training structures in the loss \eqref{eq:common_loss} contains the training cell minimizer $\bar{y}_L$, as well as perturbed states for example obtained through sub-sampling an MD trajectory, or simply rattling the atom positions.  

We now introduce {\it matching conditions} between the reference model and MLIPs that are directly motivated by such a loss function and training set but are technically more stringent, which makes a rigorous error analysis tractable.
Let $\delta>0$ and $B_{\delta}(\bar{y}_L) \subset \Us^{\rm per}_{L}$ all periodic  atomic displacements at (energy-norm) distance at most $\delta$ from $\bar{u}_L$. 
Then the matching condition for energy and forces are, respectively, defined by 
\begin{align}\label{eq:ED}
\varepsilon^{\rm E} &:= \max_{y_L \in B_{\delta}(\bar{y}_L)} \big| \E_L(y_{L}) - \E^{\rm ML}_L(y_{L})\big|, \qquad \text{and} \\ 
    \label{eq:ffit}
\varepsilon^{\rm F} &:= \max_{y_L \in B_{\delta}(\bar{y}_L)} \big\|-\nabla \E_L(y_{L})+ \nabla \E^{\rm ML}_L(y_{L})\big\|_{(\Us^{\rm per}_{L})^{*}}, 
\end{align}
where $\|\cdot\|_{(\Us^{\rm per}_{L})^{*}}$ is the dual norm of $\Us^{\rm per}_{L}$. 

Mathematically, to guarantee convergence of MLIP equilibria, one needs to ensure stability and this requires accuracy of the MLIP hessian, hence we also introduce a hessian matching condition 
\begin{align}\label{eq:FCD}
\varepsilon^{\rm H} &:= \big\|\nabla^2 \E_L(\bar{y}_{L}) - \nabla^2 \E^{\rm ML}_L(\bar{y}_{L})\big\|_{\mathcal{L}(\Us^{\rm per}_{L}, (\Us^{\rm per}_{L})^{*})} \nonumber \\
 & \qquad +  
 \big\|\nabla^2 \E_L^{\rm hom}(x_0) - \nabla^2 \E^{\rm ML, hom}_L(x_0)\big\|_{\mathcal{L}(\Us^{\rm per}_{L}, (\Us^{\rm per}_{L})^{*})}, 
\end{align}
where the second term represents the force constant error on the homogeneous lattice with the identical mapping $x_0(\ell)=\ell$ for $\ell\in\L_L$. The rationale behind establishing this matching condition stems from our stability analysis (cf.~\eqref{eq:s21a}), which necessitates control of the Hessian at defect cores as well as at the far field (homogeneous lattice).

Finally, it turns out that to accurately model the elastic fields between dislocation cores one should provide some measure of accuracy of the Cauchy--Born continuum linear elastic response. This can be understood in terms of accuracy of the stress (or, alternatively, the virial): Let $W_{\rm cb}(\mathsf{F})$ (resp. $W^{\rm ML}_{\rm cb}$) denote the unit cell energy per unit volume under deformation $\mathsf{F}$, for the reference model (resp. MLIP model) detailed in \eqref{eq:cbW}, then $\partial_{\mathsf{F}} W_{\rm cb}$ is the stress. It is common in MLIP fitting to add stresses to the loss function. For the purpose of our analysis it is convenient to introduce the matching conditions 
\begin{align}
\label{training_V}
\vfit_j := \big|\partial^{j+1}_{\mathsf{F}} \Wcb(\mathsf{I}) - \partial^{j+1}_{\mathsf{F}} \Wcb^{\rm ML}(\mathsf{I}) \big|, 
\qquad \text{for } j = 1, 2, 
\end{align}
where $\mathsf{I}$ is the identity matrix. 
The matching condition $\vfit_1$ represents the error in the continuum linear elastic response, while the the matching condition $\vfit_2$ measures a leading order error in the nonlinear elastic response. 


\subsection{Error estimates}
\label{sec:sub:error}
%
We are now in a position to quantify the geometry and energy errors committed during MLIP simulations of dislocations in terms of the fit accuracy.  The proof of the following theorem is given in the~\ref{sec:apd:proof}.

\begin{theorem}\label{them:geometry} 
Suppose that $\bar{u}$ is a strongly stable equilibrium of~\eqref{energy-difference-per} satisfying Assumption~\ref{ass:existence}. 
Then, for $\ffit, \varepsilon^{\rm H}, \vfit_1$ sufficiently small, there exists an equilibrium $\bar{y}^{\rm ML}$ of~\eqref{energy-difference-per-ML} such that
\begin{align}\label{eq:geoerr}
\|D\bar{y} - D\bar{y}^{\rm ML}\|_{\ell^2(\L_N)} &\leq~ C^{\rm G} \cdot \sqrt{n_{\D}} \cdot \big( \varepsilon^{\rm F} + \log^{1/2}(L_{\D})\cdot \vfit_1 + L^{-1}\cdot\varepsilon^{\rm V}_{2} + L^{-2}\big), \\
\label{eq:ergyerr}
\big| \E(\bar{y}) - \E^{\rm ML}(\bar{y}^{\rm ML}) \big| &\leq~ C^{\rm E} \cdot n_{\D} \cdot \Big( \big( \varepsilon^{\rm F} + \log^{1/2}(L_{\D})\cdot\vfit_1 + L^{-1}\cdot \varepsilon^{\rm V}_{2} + L^{-2}\big)^2 + L^{-2} + \varepsilon^{\rm E}  \Big),
\end{align}
where both constants $C^{\rm G}$ and $C^{\rm E}$ are independent of $N, n_{\D}$ and $L$.
\end{theorem}


\begin{remark}
    The higher-order ($j\geq 3$) derivatives of the virial do not inherently lead to a systematic improvement in convergence rates with respect to $L$, as the Cauchy--Born (continuum) modeling error ($L^{-2}$) dominates in this case. A detailed mathematical explanation can be found in the proof provided in~\ref{sec:apd:proof}. While this scenario gains importance when dealing with more intricate defects such as cracks, potential remedies may involve the training of higher-order Cauchy-Born models~\cite{wang2021priori} or incorporating higher-order boundary conditions~\cite{braun2022higher}. 
\end{remark}

\medskip

The error estimates in the foregoing theorem identify how the geometry error and the energy error depend on data-oriented approximation parameters: model accuracy on the training domain and also the size of training domain, $L$. The latter dependence was also observed in \cite{ortner2022framework} and is a somewhat unexpected effect. Regardless, our result justifies and makes precise the intuition that training a local interaction law on small training domains results in accurate predictions in large-scale complex simulations provided that local snapshots of the encountered atomic environments are contained in the training set. However, the result goes beyond this. (1) We demonstrate precisely how different observations contribute to prediction errors; and (2) we identify remaining error terms that are difficult to predict by purely geometric intuition. 

Our error estimates lead to a few immediate observations:  If we only consider energy and force in training, then the prediction errors diverge as the dislocation separation distance $L_\D \to \infty$. If we construct the {\it approximated} energy $\E^{\rm ML}$ such that the {\it matching conditions} of $\vfit_1$ and $\varepsilon_2^{\rm V}$ are exactly zero, we obtain rates of convergence in terms of $L$. Conversely, if $L$ is sufficiently large, the errors then depend on the {\it matching conditions} $\varepsilon^{\rm E}, \varepsilon^{\rm F}, \vfit_1$. These limiting cases are summarized in Table~\ref{table-e-mix}. We will see in Section~\ref{sec:sub:numerics} that these rates are indeed sharp. 

\begin{table}
	\begin{center}
	\vskip0.2cm
	\begin{tabular}{|c|c|c|c|c|}
	\hline
		Errors & {$\varepsilon^{\rm E, F}=0$} & {$\varepsilon^{\rm E, F} = \vfit_1 =0$} & {$\varepsilon^{\rm E, F} = \vfit_1 = \varepsilon_2^{\rm V} =  0$} & {$L$ sufficiently large} \\[1mm]
		\hline
		Geometry &  $\log^{1/2}(L_{\D})$ & $L^{-1}$ & $L^{-2}$ & $\varepsilon^{\rm F}+\log^{1/2}(L_{\D})\vfit_1$
		 \\[1mm]
		 \hline
		Energy &  $\log^{1/2}(L_{\D})$ & $L^{-2}$ & $L^{-2}$ & $ \varepsilon^{\rm E} + (\varepsilon^{\rm F})^2+\log(L_{\D})(\vfit_1)^2$
		\\[1mm]
		\hline 
	\end{tabular}
	\medskip 
	\caption{Limiting cases of error decay with respect to $L$ and the matching conditions $\varepsilon^{\rm E}, \varepsilon^{\rm F}$, $\varepsilon^{\rm V}_{1}$ and $\varepsilon^{\rm V}_{2}$.}
	\label{table-e-mix}
	\end{center}
\end{table}

There are further insights we can gain from Theorem~\ref{them:geometry}: First, the size of training domain can significantly affect the quality of the fitted model. Secondly, we see the importance of fitting stresses in reducing the effect of the training domain size when predictions involve long-range elastic fields.
Finally, our estimates provide a clear guidance on how energy, force and elastic constant observations should be weighted in the least squares loss function, in particular suggesting the optimal balance $\efit \approx (\ffit)^2\approx \log(L_{\D})(\vfit_1)^2$, in particular putting much higher emphasis on the energy fit, justifying another common practice in MLIPs parameter estimation.

\section{Numerical Tests}
\label{sec:numer}
The generalization analysis of the previous section gives insights into the construction of accurate MLIPs. It directs the selection of training data and the assignment of weights. We now use these results to guide a concrete MLIP implementation and test that implementation on a range of carefully designed numerical experiments to illustrate the theoretical predictions.


\subsection{Constructions of MLIPs}
\label{sec:sub:consACE}

\subsubsection{Parameterisation}
\label{sec:sub:sub:para}
First, we need to choose a parameterisation of MLIPs from an abundance of available options~\cite{Bart10, behler07, Drautz19, Shapeev16, oord19}. We choose to employ the linear atomic cluster expansion (ACE)~\cite{2019-ship1, Drautz19,  lysogorskiy2021performant} which has achieved a high accuracy comparable to state-of-the-art models~\cite{lysogorskiy2021performant} despite its relative simplicity. However, we do not consider this selection as essential and expect to obtain similar results with other models. 

Since the choice of MLIP architecture is non-essential and since the ACE model has been described in-depth in other references, we only briefly review the most salient details: Given a cutoff radius $r_{\rm cut}>0$, let ${\bf y}_\ell := \{y_{\ell m}\}_{m}$ be a collection of atom positions relative to a centre-site $\ell$, i.e., $y_{\ell m} = y(\ell) - y(m)$.
The ACE site energy $\E^{\rm ML}_{\ell}$ is written as a linear expansion
\begin{eqnarray}\label{eq:ace}
 \E^{\rm ML}_{\ell}({\bf y}_\ell; \pmb{c}) := \E^{\rm ACE}_{\ell}({\bf y}_\ell; \pmb{c}) = \sum_{B\in\pmb{B}} c_B B\big(\{y_{\ell m}\}_{|y_{\ell m}| < r_{\rm cut}}\big), 
\end{eqnarray}
where $B$ are the ACE basis functions and $\pmb{c} := \{c_B\}_{B\in\pmb{B}}$ are the parameters that will be estimated by minimizing a least squares loss. The basis functions $B$ are invariant under rotations, reflections and permutations of an atomic environment. Moreover, they are naturally body-ordered which gives a physically interpretable approximation parameter to converge the fit accuracy. A detailed review of the ACE model and its approximation parameters is provided in the~\ref{sec:ACE} and in the references~\cite{2019-ship1, Drautz19,  lysogorskiy2021performant}. The specific flavour and implementation of the ACE model that we employ is described in \cite{witt2023acepotentials}.

\subsubsection{Training sets and loss}
\label{sec:sub:sub:train_and_loss}
Following the generalisation analysis, our aim is to construct ACE models that match a reference model in the sense of making $\efit, \ffit, \varepsilon^{\rm H}, \vfit_1$ and $\varepsilon^{\rm V}_2$ small. 
The matching conditions on virial stress ($\vfit_1$ and $\varepsilon^{\rm V}_2$) can be incorporated directly into the loss. However, the matching conditions $\efit, \ffit$ and $\varepsilon^{\rm H}$ are computationally intractable, since they are specified in terms of max-norms over an infinite set of displacements. Because of this, we have to deviate slightly from our rigorous analysis setting.

We first introduce the training set, $\mathfrak{R}$, i.e. the list of training structures:  The complete neighbourhood $B_{\delta}(\bar{y}_L)$ used in the analysis is replaced with a finite number of random samples taken from $B_{\delta}(\bar{y}_L)$. 
To be precise, we first obtain $\bar{y}_L$ by solving the geometry optimization defined on the training domain. Next, given a parameter $\alpha$ and the number of the configurations in $\mathfrak{R}$ as $N_{\rm train}:= \#\mathfrak{R}$, we randomly perturb atom positions from $\bar{y}_L$ by $\alpha$ for $N_{\rm train}$ times. 
Throughout our numerical experiments, we choose two values of parameter $\alpha$ representing two levels of perturbation, i.e., $\alpha=0.5$\AA~and $\alpha=0.1$\AA. 
This completes the specification of the training set $\mathfrak{R}$. Far more sophisticated methods exist, but we aim to stay as close as possible to the setting of our analysis. 
We also produce a test set by the same method. The number of configurations in training  and test sets as $N_{\rm train}$ and $N_{\rm test}$ will be specified for each individual example.

Next, we consider the construction of a loss function inspired by our theory. We consider the same approximations of the {\it matching conditions} as those shown in~\cite{ortner2022framework}. That is, we consider the computable $\ell^2$-norm for energy and force matching and we drop the hessian matching entirely since we have found that only fitting forces and energies already results in a sufficiently good accuracy of $\varepsilon^{\rm H}$~\cite[Section 4.2]{ortner2022framework}. (So far, we have no rigorous explanation for this observation.) Given these approximations, the training set $\mathfrak{R}$ constructed above and the parameterisation defined by \eqref{eq:ace}, we determine the parameters $\{c_B\}$ by minimising the following loss function
\begin{align}\label{cost:energymix}
\mathcal{L}\big(\{c_B\}\big) := \sum_{y_R \in \mathfrak{R}}\Big(W_{\rm  E}& \big|\E_L(y_R) - \E^{\rm ML}_L(y_R; \{c_B\})\big|^2 + W_{\rm F}\big|\nabla \E_L(y_R) - \nabla \E^{\rm ML}_L(y_R; \{c_B\}) \big|^2\Big) \nonumber \\
&+ W_{\rm d1V} \cdot \vfit_1 + W_{\rm d2V} \cdot \vfit_2,
\end{align}
with additional weights $W_{\rm E}, W_{\rm F}, W_{\rm d1V}$ and $W_{\rm d2V}$.
According to \eqref{eq:ergyerr} (or Table \ref{table-e-mix}), we choose $W_{\rm E} \gg W_{\rm F} \approx W_{\rm d1V}$ in practice such that the balance $\efit \approx (\ffit)^2 \approx \log(L_{\D})(\vfit_1)^2$ can be achieved. The details will be provided for different model problems in the next section. In order to observe different convergence behaviour in terms of $L$ in practice, we test three cases that include different components of observations. Energy and force observations will always be included. To remove the virial observations we simply set some or both of $W_{\rm diV} = 0$. 

The loss function \eqref{cost:energymix} is quadratic in the parameters $\{c_B\}$ and can therefore be minimised using Bayesian linear regression schemes. In our implementation we employ the Automatic Relevance Determination (ARD) \cite{wipf2007new} to achieve the parameter estimation, which is a known statistical technique used to automatically determine the relevance of input features or variables in a predictive model.

\subsection{Numerical results}
\label{sec:sub:numerics}

In this section, we conduct several dislocation simulations to verify our theoretical analysis. As the reference model we will apply the empirical interatomic potentials instead of electronic structure models. This simplified scenario allows us to perform larger-scale simulation which is required to clearly observe the expected convergence results. 
Conceptually we expect that the results would not change if we used an electronic structure model, but it seems impossible at present to computationally verify this.
We present numerical tests for three prototypical examples:  
\begin{enumerate} 
\item {\it Screw dislocations in W: } We consider a quadrupole screw dislocation configuration in W.
An embedded atom model (EAM) \cite{Daw1984a} is applied. 

\item {\it Edge dislocations in Si: } We perform tests on a quadrupole edge dislocation in Si,
modelled by 
an optimized bond-order interatomic potential~\cite{pun2017optimized}. 

\item {\it Screw dislocations in NiAl: } We also conduct the numerical experiments on a multilattice crystal NiAl. The EAM model for NiAl is provided by the open-source interatomic potential library \textsf{OpenKIM}~\cite{tadmor2011potential}. Though our rigorous error estimate is only formulated for single-species Bravais lattices, it is conceptually straightforward to generalize it to multi-lattices. Example (3) verifies this numerically. 
\end{enumerate}

\noindent All numerical tests are implemented in open-source {\tt Julia} packages {\tt JuLIP.jl} \cite{gitJuLIP} for the implementation of molecular simulation algorithms and {\tt ACEpotentials.jl} \cite{witt2023acepotentials, gitACEpotentials} for the construction of ACE basis and the fitting of ACE models. 

\subsubsection{Quadrupole screw dislocations in W}
\label{sec:sub:sub:screw_W}
\begin{figure}[!htb]
    \centering
    \subfigure[Simulation domain $\Lambda_N$]{
    \includegraphics[height=5cm]{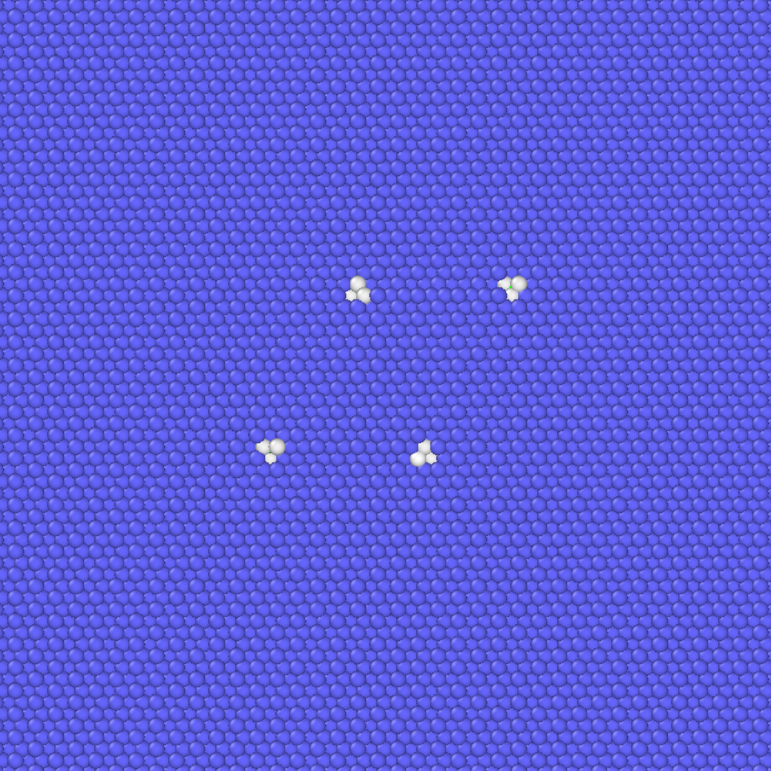}}\qquad \qquad 
    \subfigure[Training domains $\Lambda_L$]{\label{subfig:1:train}
    \includegraphics[height=5cm]{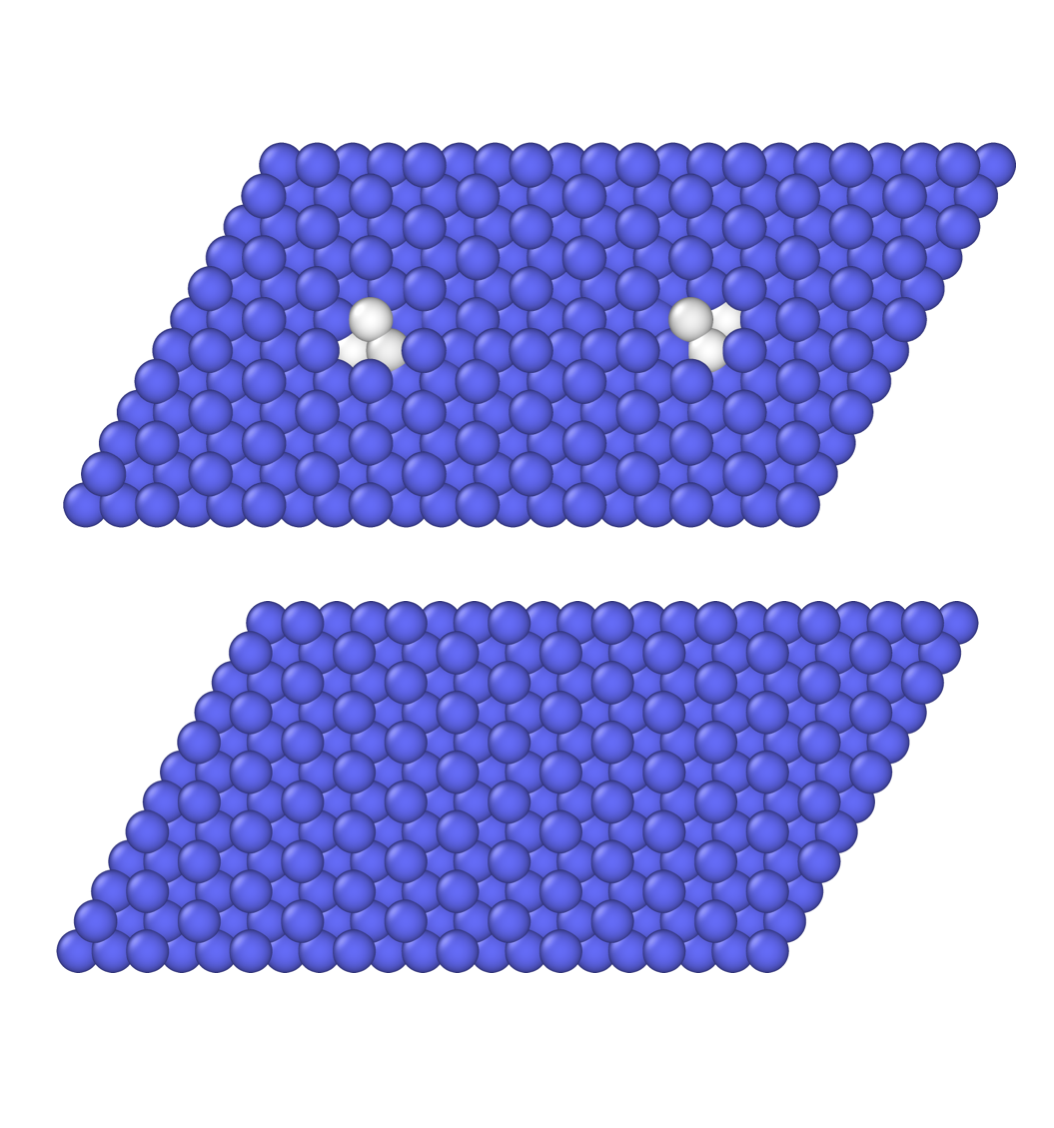}}
    \caption{Screw dislocations in W: Illustration of the simulation domain $\Lambda_N$ and the training domains $\Lambda_L$.}
    \label{fig:illu_W_screw_disloc}
\end{figure}

In this example we consider a quadruple screw dislocation in W. The simulation domain and the corresponding training domains are illustrated in Figure~\ref{fig:illu_W_screw_disloc}. The size of the simulation domain $\L_N$ is chosen to be $N=120r_0$ with $r_0$ the lattice constant of cubic solid W. Note that Figure~\ref{subfig:1:train} shows the initial state of a dipole screw dislocation instead of the equilibrium state. According to the discussion in Section~\ref{sec:sub:consACE}, we will relax these configurations first and then take random samples to construct the training set $\mathfrak{R}$.

The MLIPs are fitted by following the construction in Section~\ref{sec:sub:consACE}, where the parameters in establishing the basis functions $B$ are taken from \cite[Section 7.5]{2019-ship1}. The number of configurations in training and testing sets are set to be $N_{\rm train}=100$ and $N_{\rm test}=50$, respectively. We choose the additional weights for different kinds of observation in \eqref{cost:energymix} as $W_{\rm E}=50$, $W_{\rm F}=10, W_{\rm d1V}=10, W_{\rm d2V}=1$ in order to balance the matching conditions $\efit \approx (\ffit)^2 \approx \log(L_{\D})(\vfit_1)^2$ (cf. Table~\ref{table-e-mix}). 

\begin{figure}[!htb]
    \centering
    \subfigure[Geometry error]{
    \includegraphics[height=5.5cm]{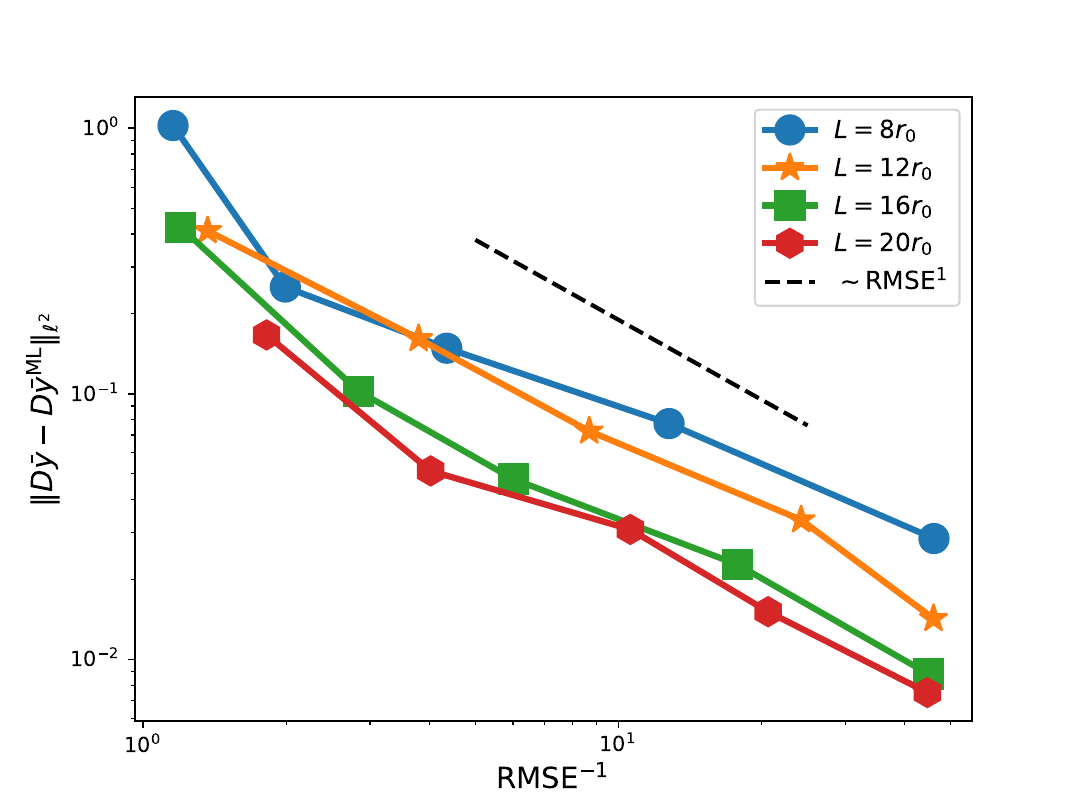}}
    \subfigure[Energy error]{
    \includegraphics[height=5.5cm]{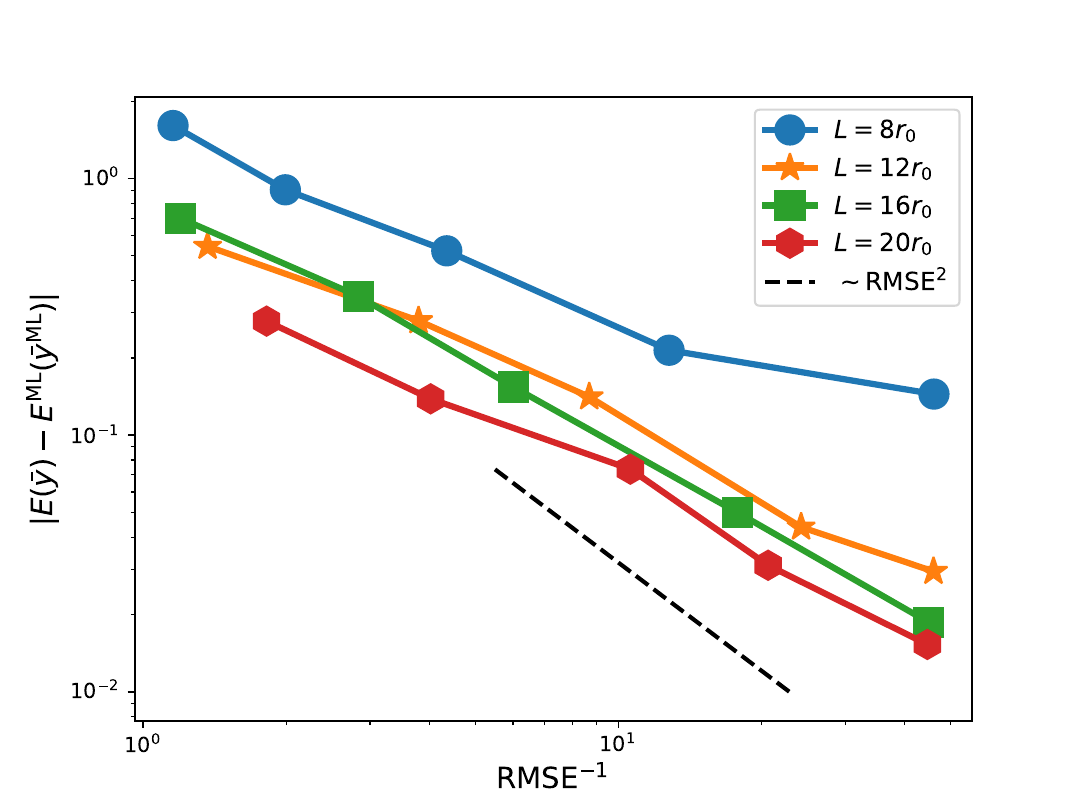}}
    \caption{Screw dislocations in W: Geometry error and energy error v.s. RMSE.}
    \label{fig:W_screw_errorvsrmse}
\end{figure}

We first test the convergence of the geometry error $\|D\bar{y} - D\bar{y}^{\rm ML}\|_{\ell^2}$ and the error in energy $|\E(\bar{y})-\E^{\rm ML}(\bar{y}^{\rm ML})|$ with respect to the root mean square error (RMSE) of test set. We choose the case that all the observations discussed in this work are taken into consideration. Figure \ref{fig:W_screw_errorvsrmse} shows that, for different size of training domain $L$, the error curves of geometry error and error in energy decrease near linearly and quadratically respectively as RMSE decreases, which reasonably matches our theoretical predictions from Theorem \ref{them:geometry}. Note that too close a match cannot be expected due to our slight departure from the rigorous framework. 

\begin{figure}[!htb]
    \centering
    \subfigure[Geometry error]{
    \includegraphics[height=5.5cm]{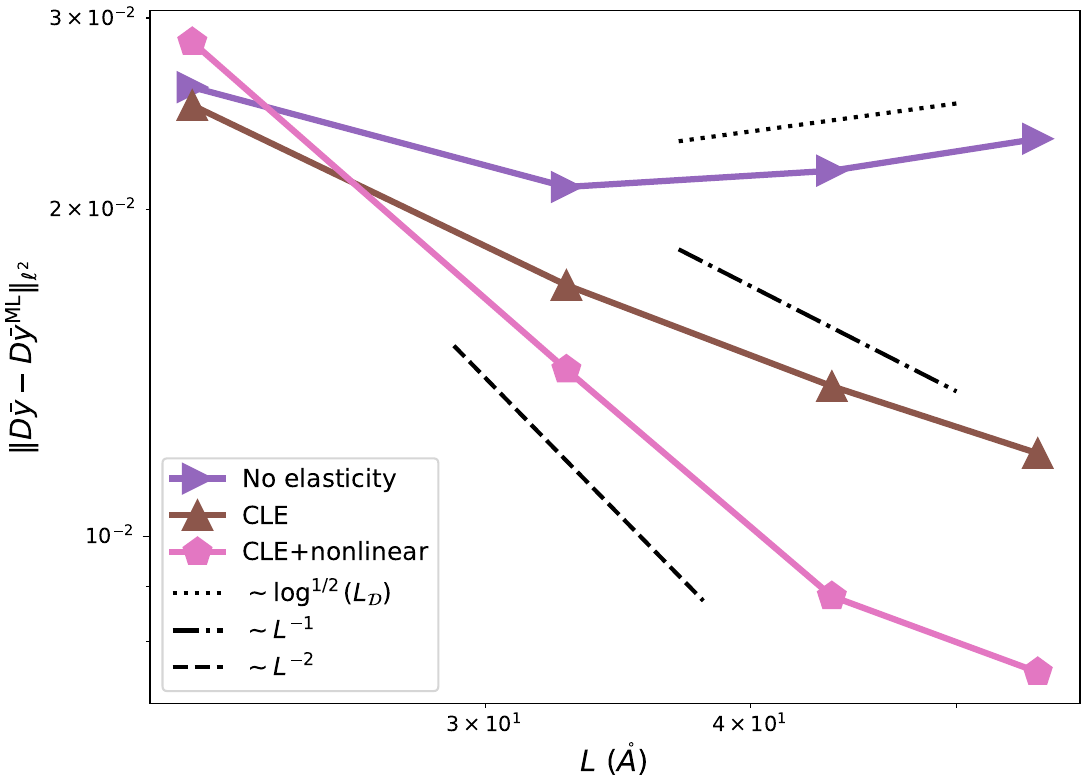}}
    \subfigure[Energy error]{
    \includegraphics[height=5.5cm]{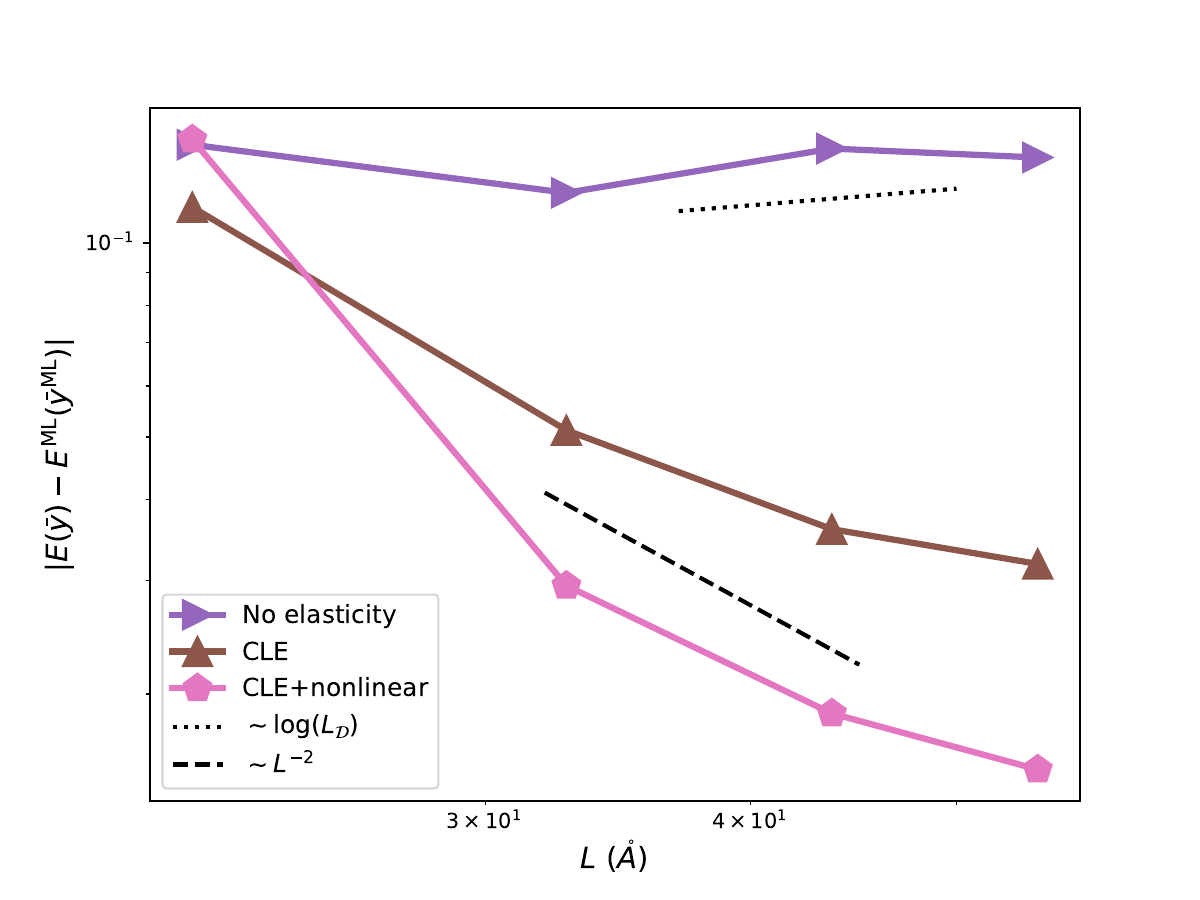}}
    \caption{Screw dislocations in W: Geometry error and energy error v.s. $L$. ``No elasticity" indicates that both $W_{\rm d1V}$ and $W_{\rm d2V}$ are equal to zero. ``CLE" signifies that $W_{\rm d2V}$ is set to zero, while ``CLE+nonlinear" encompasses all observations, including nonlinear elasticity.} 
    \label{fig:W_screw_errorvsL}
\end{figure}

Figure~\ref{fig:W_screw_errorvsL} plots the decay of geometry error $\|D\bar{y} - D\bar{y}^{\rm ML}\|_{\ell^2}$ and energy error $|\E(\bar{y})-\E^{\rm ML}(\bar{y}^{\rm ML})|$ against the size of training domain $L$. In order to observe different convergent behaviours in terms of $L$, we test three
cases that include different components of observations. The observed convergence rates in our simulations align precisely with the theoretical predictions from Theorem~\ref{them:geometry} and Table \ref{table-e-mix} for screw dislocations in W. This finding confirms the fundamental importance of training MLIPs with elastic constants in dislocation simulations, distinguishing them from simulations involving point defects~\cite{ortner2022framework}. Our results are also consistent with the best practices recommended in the MLIPs literature for dislocation simulations~\cite{grigorev2023calculation, 2021-qmmm3}.

\subsubsection{Quadrupole edge dislocations in Si}
\label{sec:sub:sub:edge-Si}

\begin{figure}[!htb]
    \centering
    \subfigure[Simulation domain $\Lambda_N$]{
    \includegraphics[height=5cm]{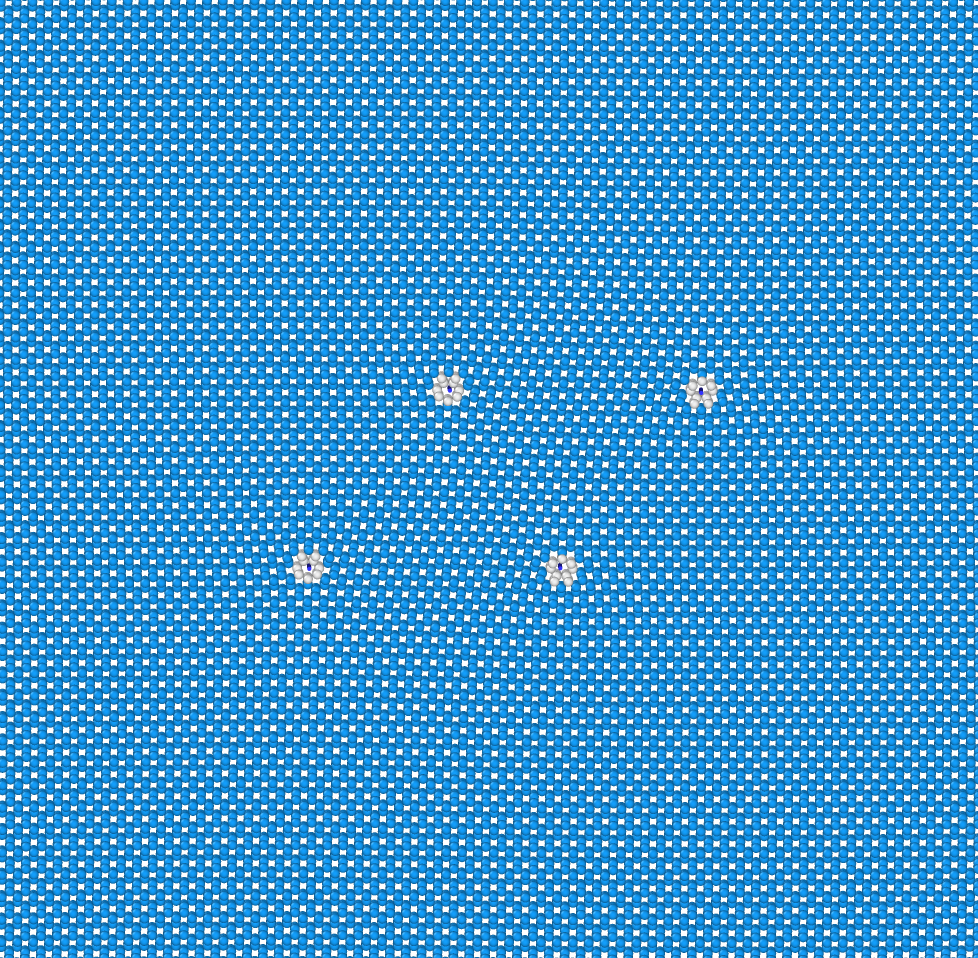}}\qquad \qquad 
    \subfigure[Training domains $\Lambda_L$]{
    \includegraphics[height=5cm]{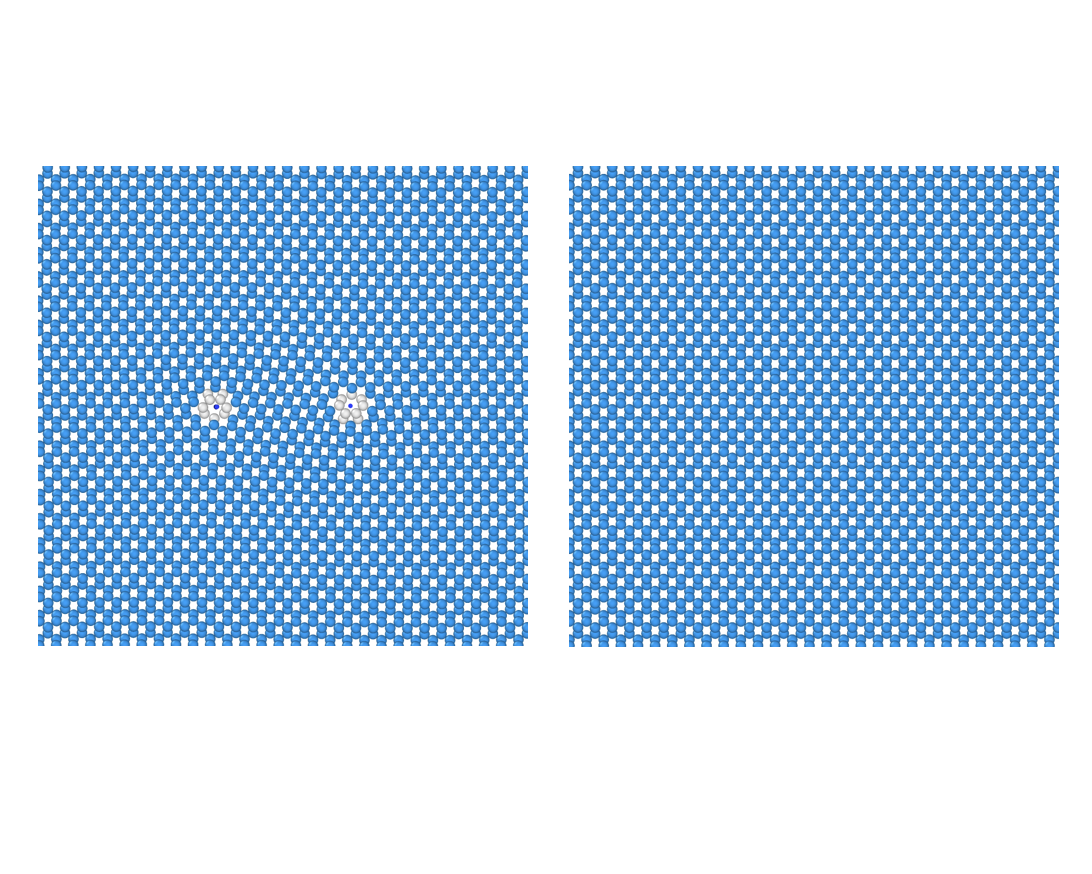}}
    \caption{Edge dislocations in Si: Illustration of the simulation domain $\Lambda_N$ and the training domains $\Lambda_L$.}
    \label{fig:illu_Si_edge_disloc}
\end{figure}

In this case we consider edge dislocations in Si. The simulation domain and the corresponding training domains are illustrated in Figure~\ref{fig:illu_Si_edge_disloc}. The size of the simulation domain $\L_N$ is chosen to be $N=100r_0$ with $r_0$ the lattice constant of solid diamond Si. The parameters in establishing the basis functions $B$ for Si are also taken from \cite[Section 7.5]{2019-ship1}. We take the same number of  configurations in training and testing sets as those in screw dislocations shown in the previous section. The additional weights are chosen to be $W_{\rm E}=100$, $W_{\rm F}=20, W_{\rm d1V}=20, W_{\rm d2V}=1$. 

\begin{figure}[!htb]
    \centering
    \subfigure[Geometry error]{
    \includegraphics[height=5.5cm]{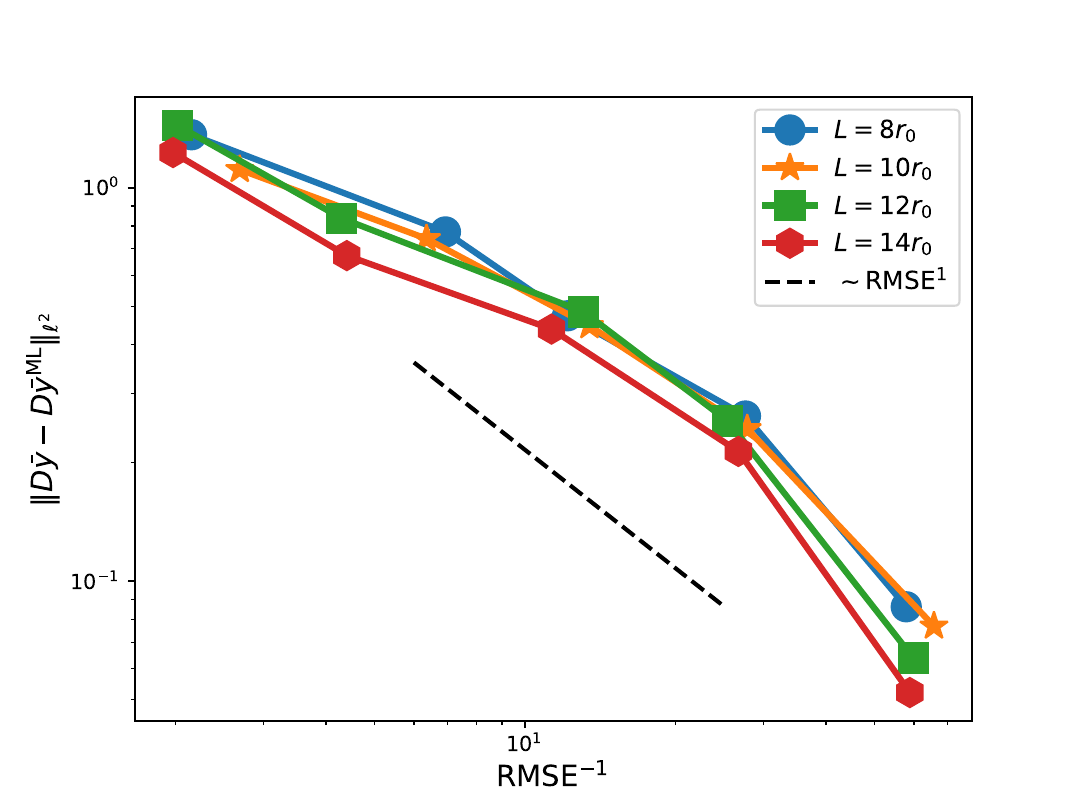}}
    \subfigure[Energy error]{
    \includegraphics[height=5.5cm]{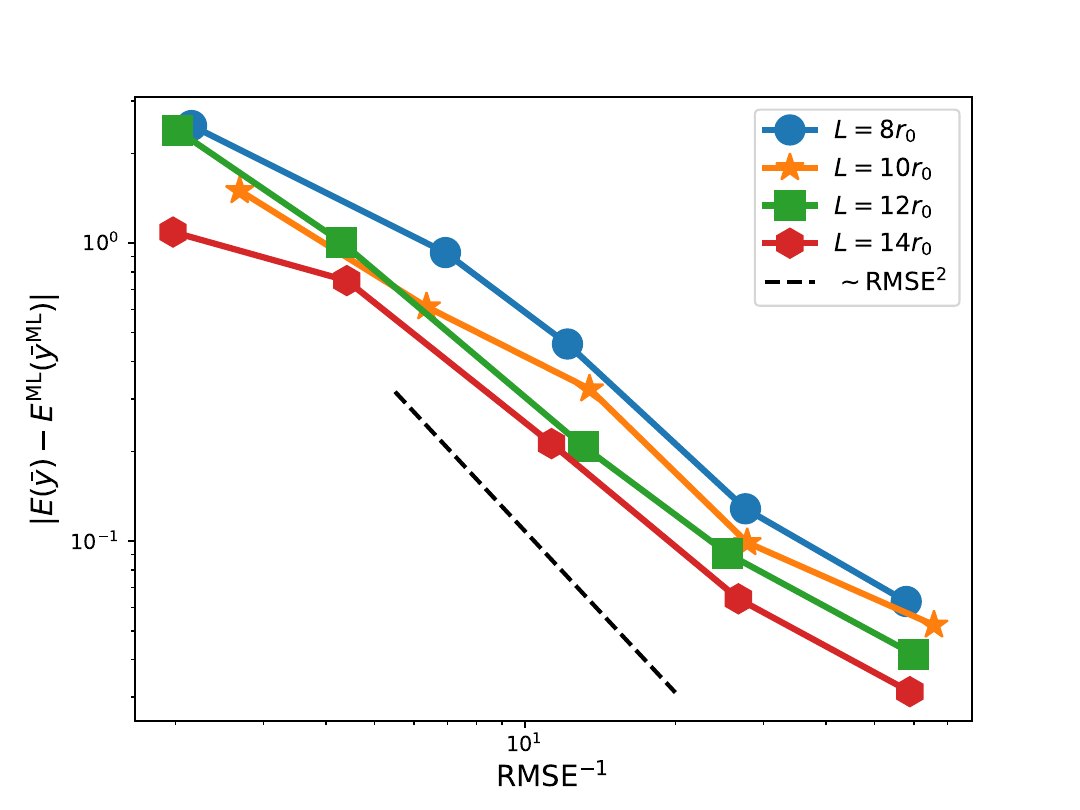}}
    \caption{Edge dislocations in Si: Geometry error and enegry error v.s. RMSE.}
    \label{fig:Si_edge_errorvsrmse}
\end{figure}

\begin{figure}[!htb]
    \centering
    \subfigure[Geometry error]{
    \includegraphics[height=5.5cm]{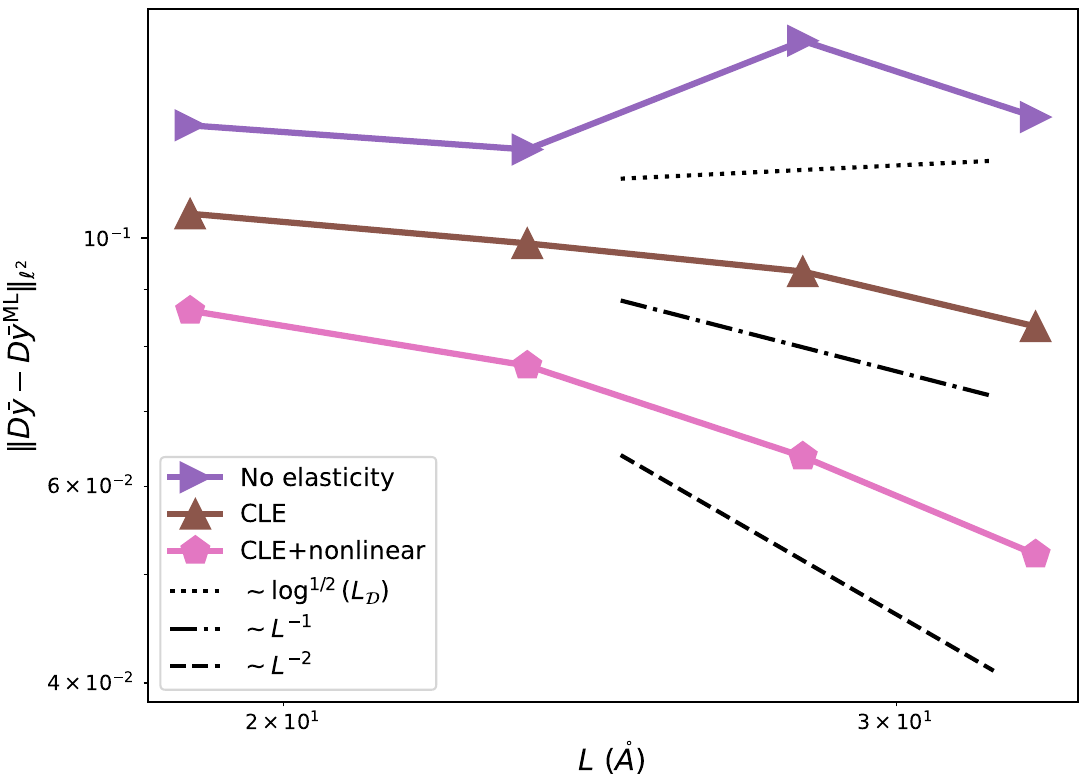}}
    \subfigure[Energy error]{
    \includegraphics[height=5.5cm]{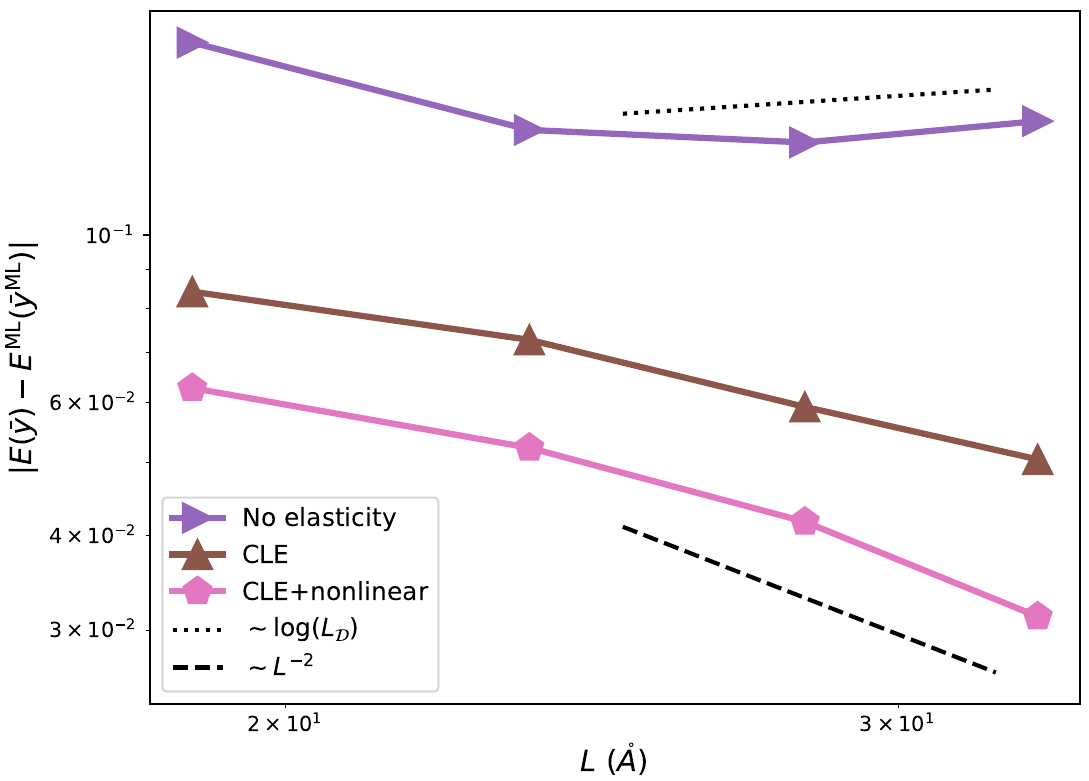}}
    \caption{Edge dislocations in Si: Geometry error and energy error v.s. $L$. ``No elasticity'' indicates that both $W_{\rm d1V} = W_{\rm d2V} = 0$. ``CLE'' signifies that $W_{\rm d2V} = 0$, while ``CLE+nonlinear'' encompasses all observations.}
    \label{fig:Si_edge_errorvsL}
\end{figure}

The convergence of geometry error and energy error against RMSE is shown in Figure~\ref{fig:Si_edge_errorvsrmse}, where the predicted convergence is again observed for this edge dislocation case. Figure~\ref{fig:Si_edge_errorvsL} plots the decay of geometry error $\|D\bar{y} - D\bar{y}^{\rm ML}\|_{\ell^2}$ and energy error $|\E(\bar{y})-\E^{\rm ML}(\bar{y}^{\rm ML})|$ against the size of training domain $L$. Our observations reveal that the convergence rates for edge dislocations align well with the theoretical predictions derived from Theorem~\ref{them:geometry} and Table \ref{table-e-mix}. This numerical verification confirms the generalization analysis for edge dislocations.

\subsubsection{Quadrupole screw dislocations in NiAl}
\label{sec:sub:sub:NiAl-screw}

In this example, we extend our dislocation simulations to multilattice crystals. The theoretical analysis of accurately modeling the geometric equilibrium of dislocations in multilattice systems has been previously investigated in~\cite{olson2019theoretical}. Leveraging this theoretical foundation, our generalization analysis could be applied to dislocation simulations in multilattices. To verify the effectiveness of our approach, we will conduct numerical experiments as described below. The total number of the configurations in the training and testing sets and the additional weights in \eqref{cost:energymix} are chosen to be the same as those in screw dislocations in W presented in the previous section. 

\begin{figure}[!htb]
    \centering
    \includegraphics[height=5cm]{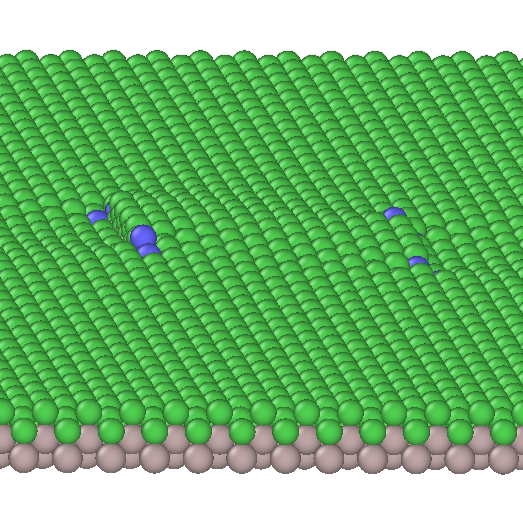}\quad
    \includegraphics[height=4.8cm]{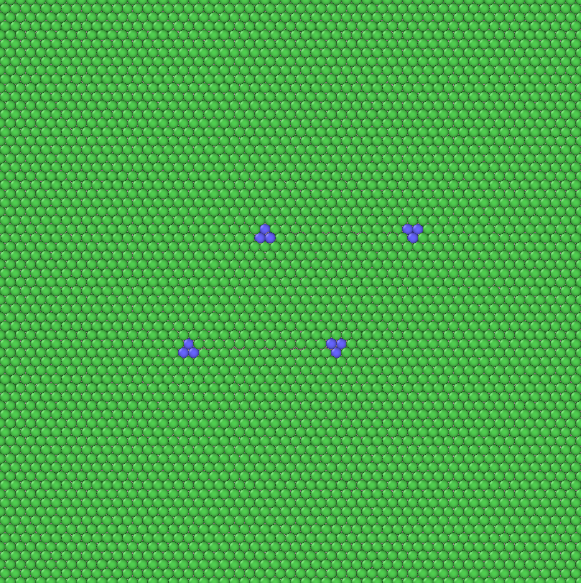}\quad
    \includegraphics[height=5cm]{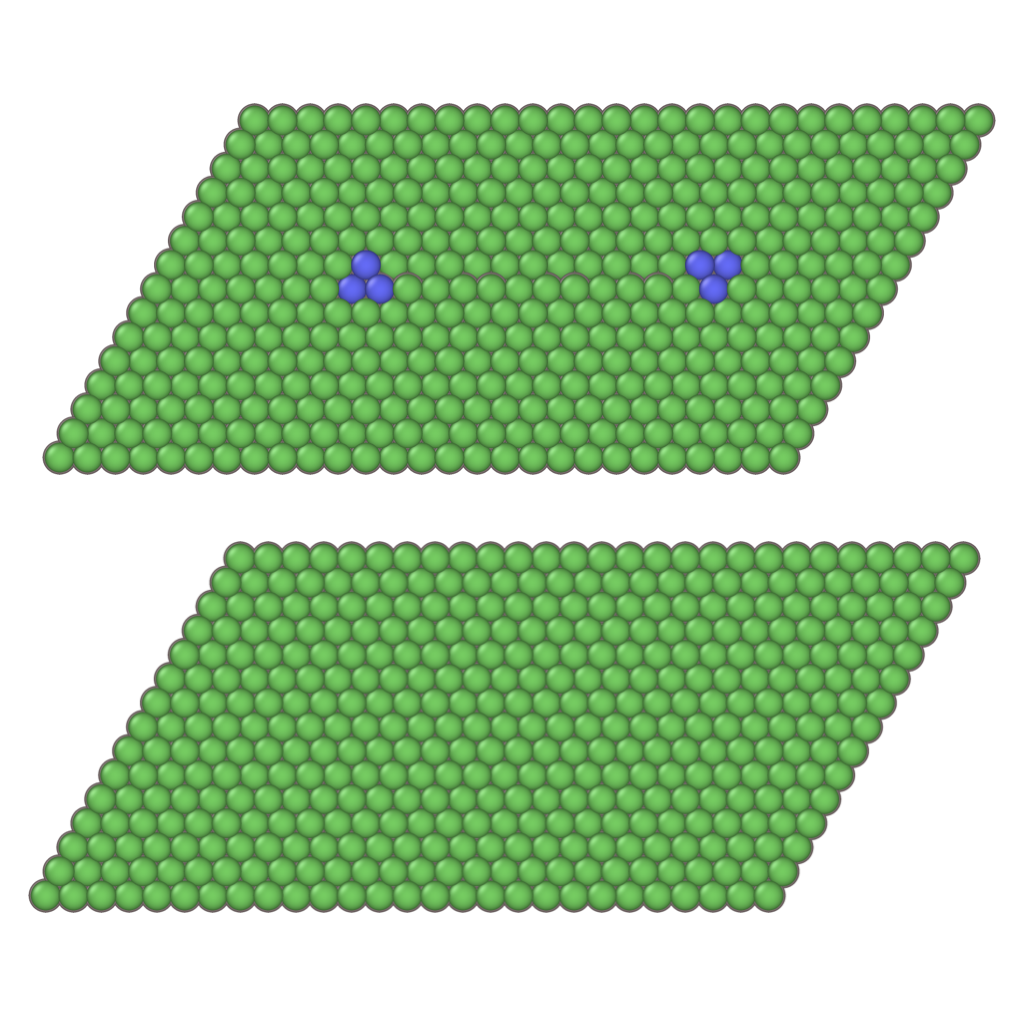}
    \caption{Screw dislocations in NiAl: Illustration of orthogonal (left) and top (middle) views of simulation domain and training domains (right).}
    \label{fig:illu_NiAl_screw_disloc}
\end{figure}

\begin{figure}[!htb]
    \centering
    \subfigure[Geometry]{
    \includegraphics[height=5.5cm]{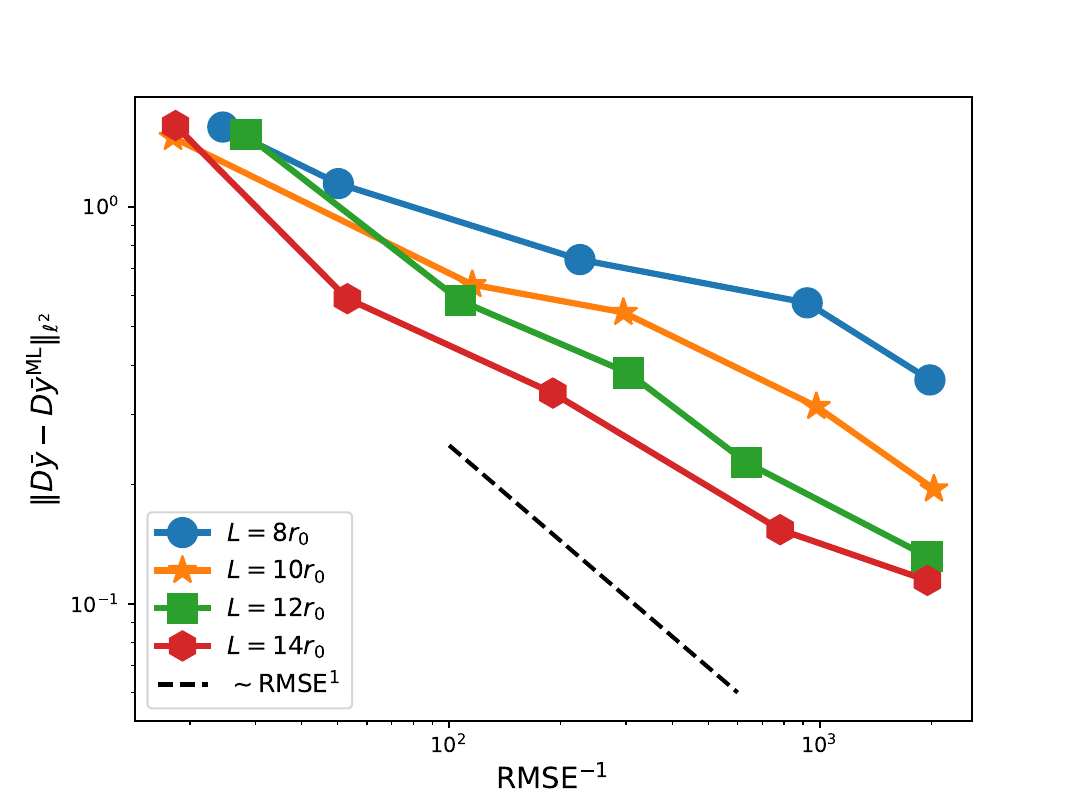}}
    \subfigure[Energy error]{
    \includegraphics[height=5.5cm]{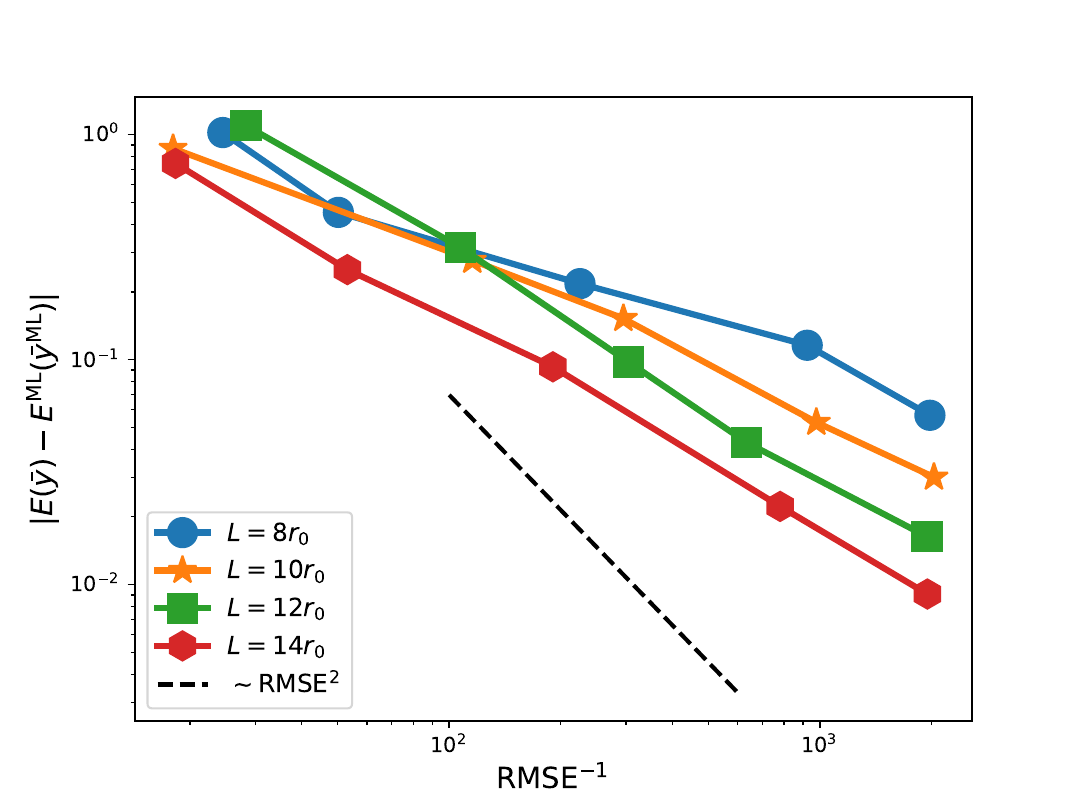}}
    \caption{Screw dislocations in NiAl: Geometry error and energy error v.s. RMSE.}
    \label{fig:NiAl_screw_errorvsrmse}
\end{figure}

\begin{figure}[!htb]
    \centering
    \subfigure[Geometry error]{
    \includegraphics[height=5.5cm]{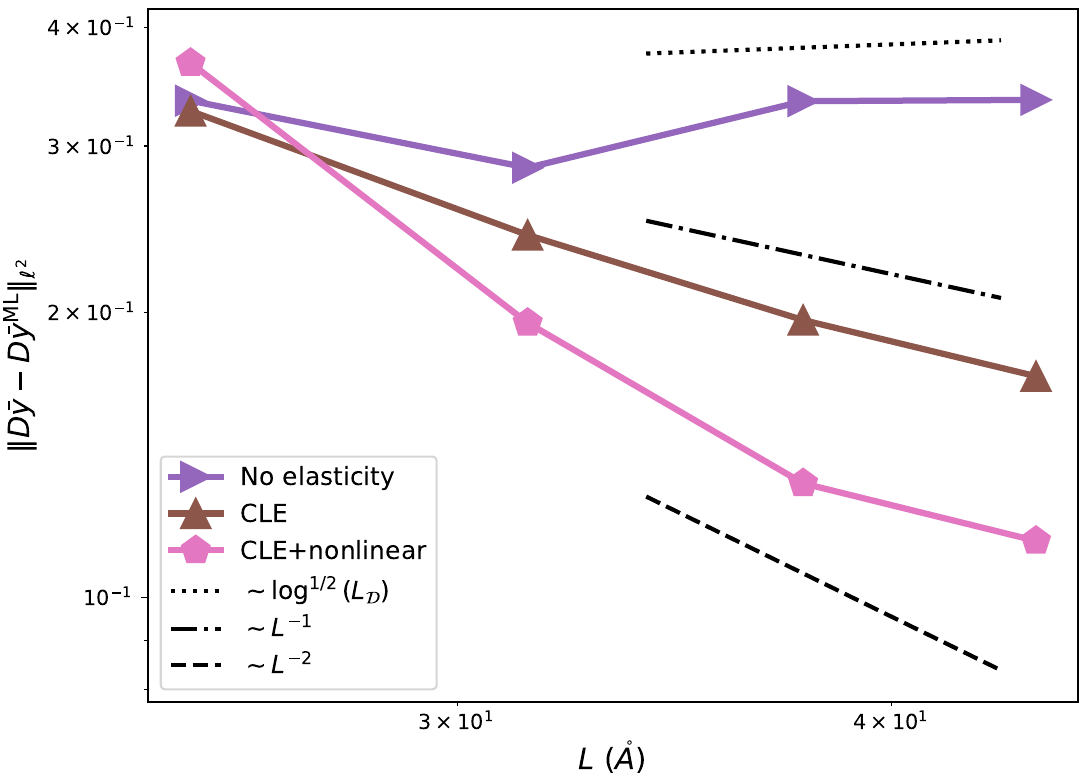}}
    \subfigure[Energy error]{
    \includegraphics[height=5.5cm]{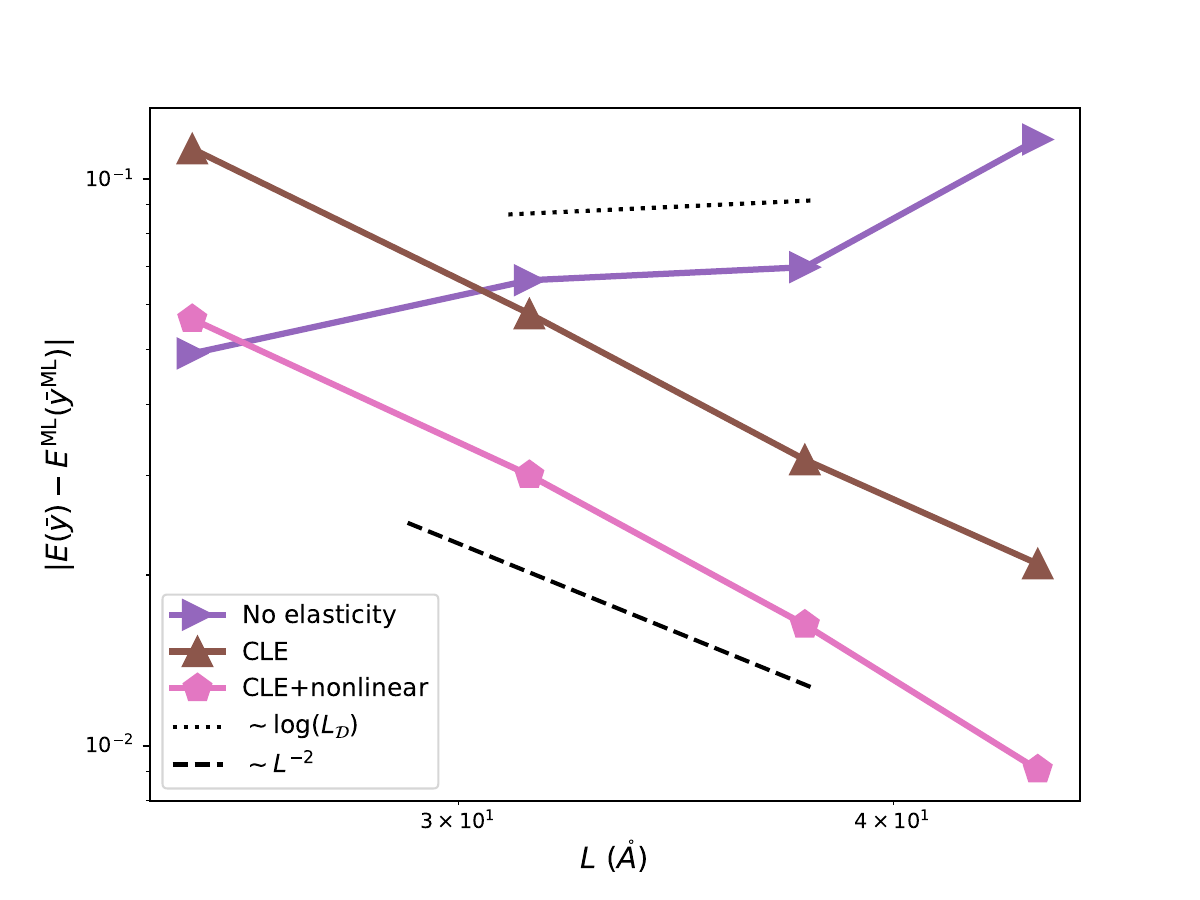}}
    \caption{Screw dislocations in NiAl: Geometry error and energy error v.s. $L$. ``No elasticity'' indicates that both $W_{\rm d1V} = W_{\rm d2V} = 0$. ``CLE'' signifies that $W_{\rm d2V} = 0$, while ``CLE+nonlinear'' encompasses all observations. }
    \label{fig:NiAl_screw_errorvsL}
\end{figure}

The convergence of geometry error and energy error against RMSE is shown in Figure~\ref{fig:NiAl_screw_errorvsrmse}. Figure~\ref{fig:NiAl_screw_errorvsL} plots the decay of geometry error $\|D\bar{y} - D\bar{y}^{\rm ML}\|_{\ell^2}$ and energy error $|\E(\bar{y})-\E^{\rm ML}(\bar{y}^{\rm ML})|$ against the size of training domain $L$. We observe that the convergence rates in our numerical experiments again align reasonably well with the theoretical predictions, providing evidence for the validity of our rigorous analysis in the context of multilattice crystals. 

\subsection{Extension: a dislocation loop}
\label{sec:sub:ext}

In this section, we explore the application of our methodologies to a dislocation loop that require a greater extent of extrapolation compared to straight dislocations. Our theory cannot be readily extended to this case, but we still expect that some of our theoretical observations can be transferred at least experimentally. 

\begin{figure}[!htb]
    \centering
    \subfigure[Simulation domain $\Lambda_N$]{
    \includegraphics[height=5cm]{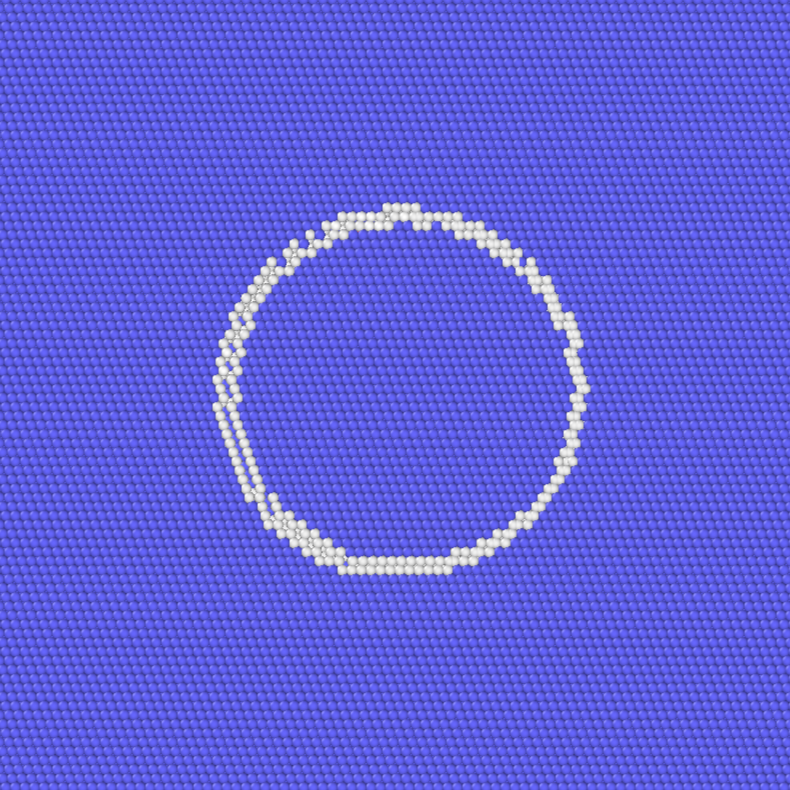}}\qquad \qquad 
    \subfigure[Training domains $\Lambda_L$]{
    \includegraphics[height=5cm]{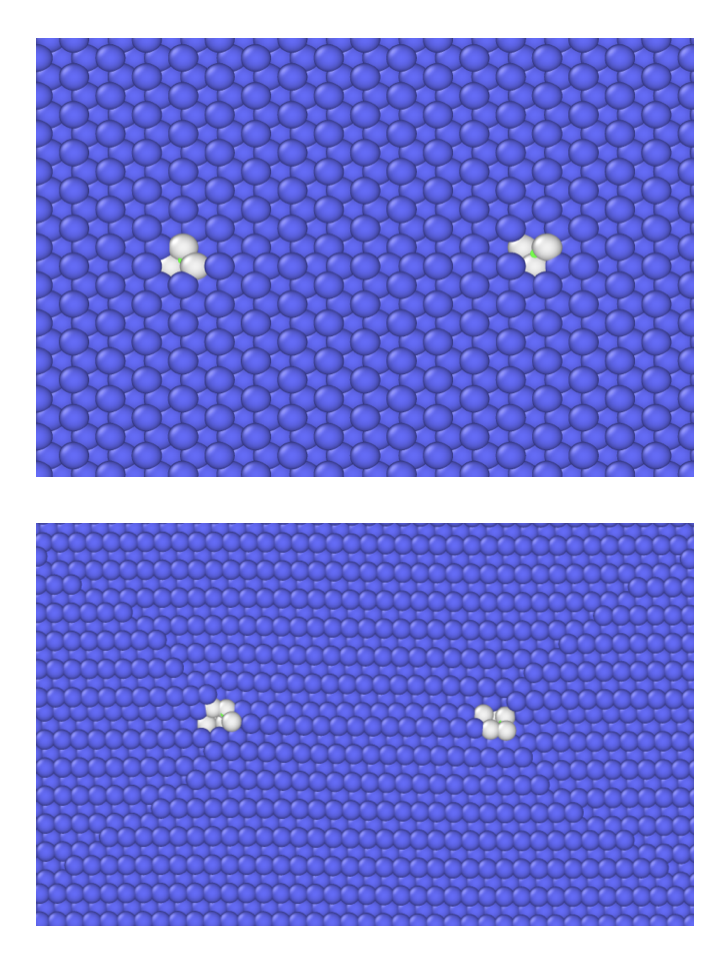}}
    \caption{A dislocation loop in W: Illustration of the simulation domain $\Lambda_N$ and the training domains $\Lambda_L$.}
    \label{fig:illu_W_disloc_loop}
\end{figure}

The simulation domain now contains an entire dislocation loop while the training domains contain a dipole screw and a dipole edge dislocation, which are illustrated in Figure~\ref{fig:illu_W_disloc_loop}. The size of the simulation domain $\L_N$ is chosen to be $N=150r_0$ with $r_0$ the lattice constant of cubic solid W. With a little abuse of notations, the diameter of dislocation loop is also denoted by $L_{\D}$. The size of training domain is taken to be $L=L_{\D}$ in practice. 

The number of configurations in training and testing sets are set to be $N_{\rm train}=150$ and $N_{\rm test}=50$, respectively. We choose the additional weights for different kinds of observation in \eqref{cost:energymix} as $W_{\rm E}=100$, $W_{\rm F}=20, W_{\rm d1V}=20, W_{\rm d2V}=1$ in order to match the optimal error balance. 

\begin{figure}[!htb]
    \centering
    \subfigure[Geometry error]{\includegraphics[height=5.5cm]{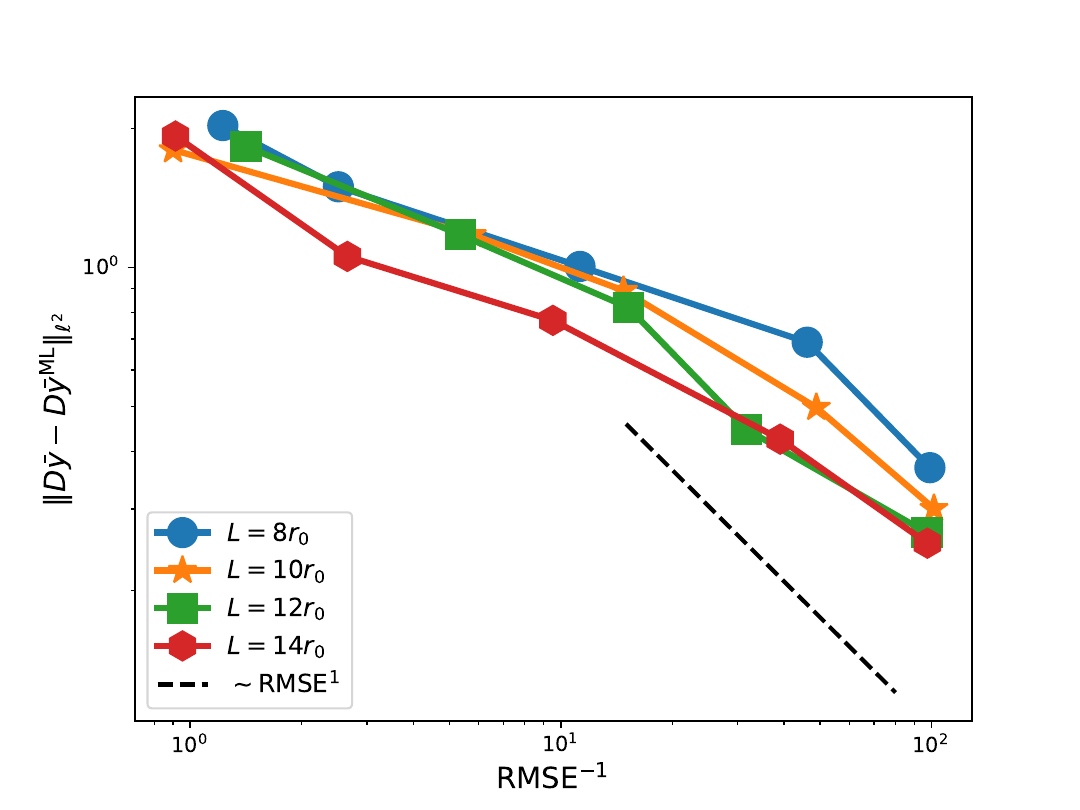}}~~\quad
    \subfigure[Error in energy]{\includegraphics[height=5.5cm]{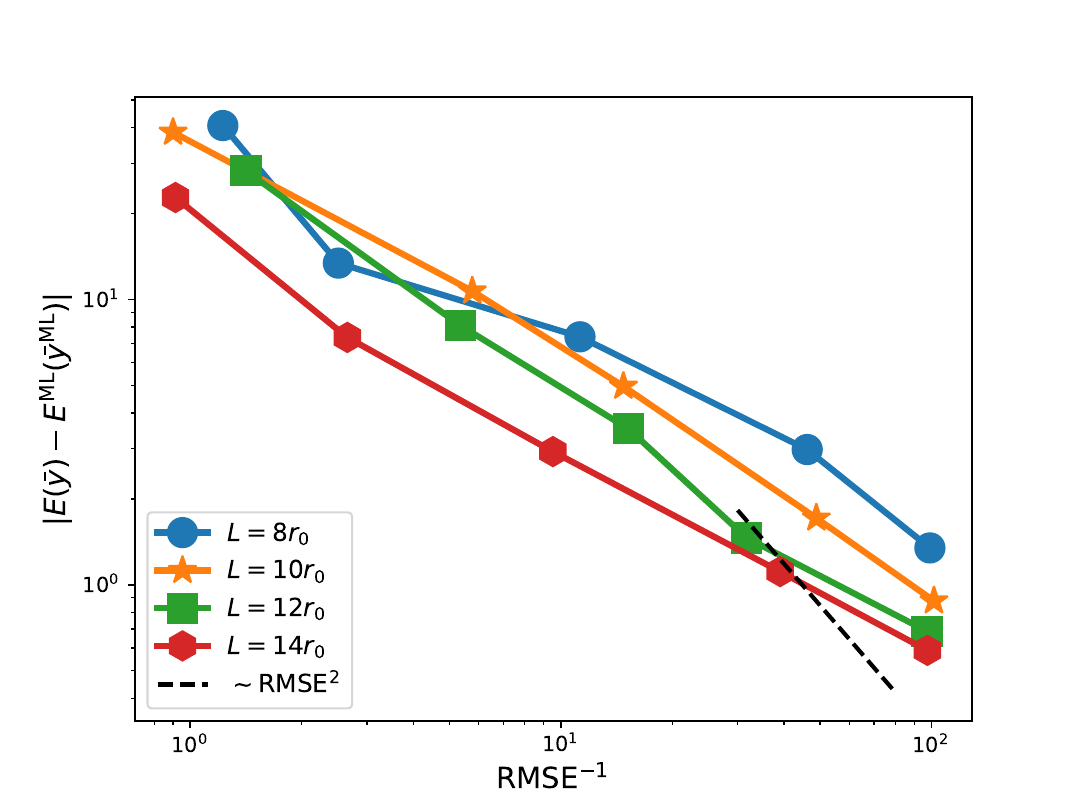}}
    \caption{A dislocation loop in W: Geometry error and energy error v.s. RMSE.}
    \label{fig:loop_errorvsrmse}
\end{figure}

\begin{figure}[!htb]
    \centering
    \includegraphics[height=5.5cm]{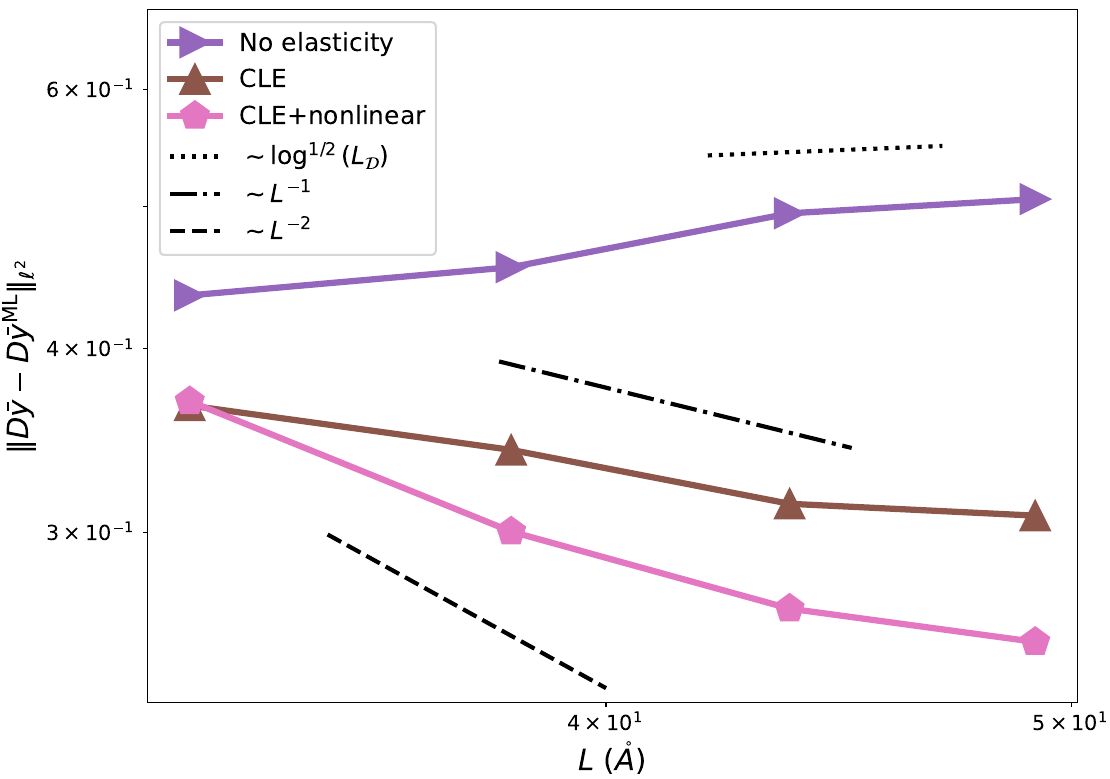}
    \includegraphics[height=5.5cm]{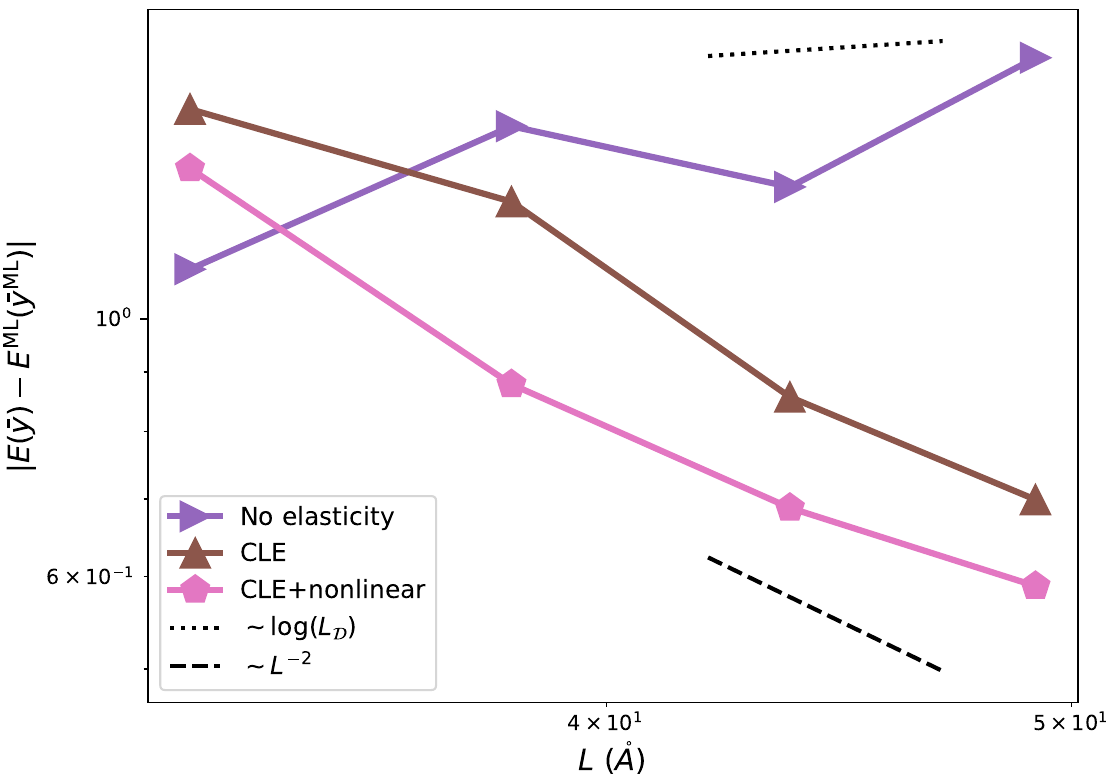}
    \caption{A dislocation loop in W: Geometry error and energy error v.s. $L$. ``No elasticity'' indicates that both $W_{\rm d1V} = W_{\rm d2V} = 0$. ``CLE'' signifies that $W_{\rm d2V} = 0$, while ``CLE+nonlinear'' encompasses all observations. }
    \label{fig:W_loop_errorvsL}
\end{figure}

The convergence of geometry error and energy error against RMSE for dislocation loop is shown in Figure~\ref{fig:loop_errorvsrmse}, where the predicted convergence is again observed for this electronic structure model. Figure~\ref{fig:W_loop_errorvsL} plots the decay of geometry error $\|D\bar{y} - D\bar{y}^{\rm ML}\|_{\ell^2}$ and energy error $|\E(\bar{y})-\E^{\rm ML}(\bar{y}^{\rm ML})|$ against the size of training domain $L$. We observe that the convergence rates of our simulations for dislocation loops surprisingly align with the theoretical predictions presented in Theorem~\ref{them:geometry} and Table \ref{table-e-mix}. This indicates that our methodologies continue to show promising results even in the case of dislocation loops. However, it is important to note that these simulations utilize empirical potentials, which can be considered as a ``low-dimensional" force-field. In the context of extending our methodologies to electronic structure models, more careful and detailed investigations are required. Specifically, the interactions between dislocations need to be taken into account and thoroughly examined. Such investigations will provide valuable insights and enable the development of more accurate and robust models for dislocation simulations in electronic structure frameworks.

\section{Conclusion}
\label{sec:conclu}
We presented a theoretical and numerical study of the generalization capability of MLIPs for dislocation simulations in a crystalline solid. We investigated the error propagation from fitting MLIPs on a small training domain to making predictions on a large simulation domain. Our analysis identifies what observations one should acquire from the reference model to obtain accurate predictions in this case. 
Our theoretical results partially justify existing best practices in the MLIPs literature, but also provide more fine-grained qualitative information about how prediction accuracy depends on the choice of training data. 
This approach also suggests a new perspective on how to approach the collection of training data, augmenting e.g. the emerging active learning approaches, and in particular also on the design of loss functions. 

The present work intends to highlight the potential of our approach in a relatively simple setting where a detailed and rigorous study is possible, but as a result is limited in scope. A large number of possible generalisations are possible, for example to more complex material and defect geometries, or how our analytic approach to error estimation might be combined with more statistical approached.

\appendix

\section{Proofs}
\label{sec:appendix:proof}

In this section, we begin by introducing the necessary concepts and models within the primary context of this work. We accomplish this by conducting a comprehensive review of the framework proposed in~\cite{chen19, Ehrlacher16, 2014-dislift}, while also adapting their approaches to align with the specific objectives of our current work. We provide a sketch of the proof of Assumption~\ref{ass:existence} for a special case in~\ref{sec:apd:as}. Building upon this, we subsequently provide the proof of generalization analysis (cf. Theorem~\ref{them:geometry}) in \ref{sec:apd:proof}.

\subsection{Preliminaries}
\label{sec:apd:pre}

A rigorous framework for modelling the geometric equilibrium of crystalline defects has been developed in~\cite{chen18, chen19, Ehrlacher16, co2020}. These works formulate the equilibration of a single crystalline defect as a variational problem in a discrete energy space that is analogous to the classical (homogeneous) Sobolev space $\dot{H}^1$. In this section, we will review the framework and adapt it to the case of multiple dislocations considered in this work along the lines of~\cite{ortner2022framework, 2014-dislift}. 

As discussed in Section~\ref{sec:sub:supercell}, a general deformed configuration of the periodically repeated lattice $\L_N^{\rm per}$, with multiple straight dislocations, is a map $y: \L_N^{\rm per}\rightarrow\R^3$, decomposed into
\[
y(\ell) = \ell + u^{\rm CLE}_{\rm per}(\ell) + u(\ell) = y^{\rm CLE}_{\rm per}(\ell) + u(\ell),
\]
where the {\it displacement} field $u: \L^{\rm per}_N\rightarrow\R^3$ is a {\it core corrector}. 

We introduce the finite difference stencil to represent the local atomistic environment. For $\ell\in\L_N^{\rm per}$ and $\rho\in\L_N^{\rm per}-\ell$, we define the finite difference
$D_\rho v(\ell) := v(\ell+\rho) - v(\ell)$. For a subset $\Rl \subset \L_N^{\rm per}-\ell$, we
define $D_{\Rl} v(\ell) := (D_\rho v(\ell))_{\rho\in\Rl}$, and we denote $Dv(\ell) := D_{\L_N^{\rm per}-\ell} v(\ell)$. For a stencil $Dv(\ell)$, we define the stencil norms
\begin{align}\label{eq: nn norm}
\big|Dv(\ell)\big|_{\mathcal{N}} := \bigg( \sum_{\rho\in \mathcal{N}(\ell) - \ell} \big|D_\rho v(\ell)\big|^2 \bigg)^{1/2}  
\quad{\rm and}\quad
\|Dv\|_{\ell^2(\L_N)} := \bigg( \sum_{\ell \in \L_N} |Dv(\ell)|_{\mathcal{N}}^2 \bigg)^{1/2},
\end{align}
where the nearest neighbours set $\mathcal{N}(\ell)$ is defined as 
\begin{align}\label{def1:Nl} 
\mathcal{N}(\ell) :=& \left\{ \, m \in \L^{\rm per}_N \setminus \ell~\Big|~\exists \, a \in \mathbb{R}^{d} \text{ s.t. }
|a - \ell| = |a - m| \leq |a - k| \quad \forall \, k \in \L^{\rm per}_N \, \right\}.
\end{align}

Next, we introduce the concept of site potential, a crucial factor that provides insight into the local energy contributions from specific atomic sites. This element holds significant importance in defining the fundamental physical model utilized in our simulations. Although quantum mechanical models often rely on total energies, there are situations where it is feasible and relevant to construct site energies~\cite{chen16, finnis03}.

The reference site potential is a mapping $V: (\mathbb{R}^3)^{\L\setminus0} \to \mathbb{R}$. 
In this paper, we make the following assumptions on the regularity and locality of the site potentials, which has been justified for some basic quantum mechanic models \cite{chen18, co2020, chen16, chen19tb}. We refer to \cite[Section 2.3 and Section 4]{chen19} for discussions of more general site potentials.
\begin{itemize}
	\label{as:SE:pr}
	\item[\assERL]
	{\it Regularity and locality:}
	For all $\ell \in \L$, $V_{\ell}\big(Dy(\ell)\big)$ possesses partial derivatives up to $\mathfrak{n}$-th order with $\n\geq 3$. For $j=1,\ldots,\n$, there exist constants $C_j$ and $\eta_j$ such that
	\begin{eqnarray}
	\label{eq:Vloc}
	\big|V_{\ell,{\bm \rho}}\big(Dy(\ell)\big)\big|  \leq
	C_j \exp\Big(-\eta_j\sum^j_{l=1}|{\bm \rho}_l|\Big)
	\end{eqnarray} 
	for all $\ell \in \L$ and ${\bm \rho} \in (\L - \ell)^{j}$. 
\end{itemize}

Although we defined the site potentials on infinite stencils $(\R^m)^{\L-\ell}$, the setting also applies to finite systems or to finite range interactions. 
It is only necessary to assume in this case that the potential $V_{\ell}(\pmb{g})$ does not depend on the reference sites $\pmb{g}_{\rho}$ outside the interaction range.

The energy-difference functional is then defined by 
\begin{eqnarray}\label{energy-difference-per}
\E(y) := 
\sum_{\ell\in\L_N} V_{\ell}\big(Dy(\ell)\big).
\end{eqnarray}

An equilibrium defect geometry is obtained by solving
\begin{align}\label{eq:variational-problem-per-apd}
    \bar{y} \in \arg\min \big\{ \E(y),~ 
    y-y^{\rm CLE}_{\rm per} \in \Us_N^{\per} \big\}.
\end{align}

In order to derive the {\it matching conditions} for virial stress, we briefly review the {\it Cauchy-Born rule}~\cite{weinan2007cauchy, co2013}, which relates the movement of atoms in a crystal to the overall deformation of the bulk solid.
For $\mathsf{F} \in \R^{3\times 3}$, the Cauchy-Born rule makes an approximation such that in a crystalline solid subject to a small strain, the positions of the atoms within the crystal lattice follow the overall strain of the medium. More precisely, the reference and MLIPs site potentials are approximated by the Cauchy-Born elastic energy density functional $W_{\rm cb}, ~ W^{\rm ML}_{\rm cb} : \R^{3\times 3} \rightarrow \R$ respectively, with
\begin{eqnarray}
\label{eq:cbW}
W_{\rm cb}(\mathsf{F}) := V(\mathsf{F}\cdot\L_{*}) \qquad \text{and} \qquad W^{\rm ML}_{\rm cb}(\mathsf{F}) := V^{\rm ML}(\mathsf{F}\cdot\L_{*}),
\end{eqnarray}
where $\L_{*}:=\L\setminus 0$. The derivative (virial stress) and even higher order derivatives with respect to the deformation $\mathsf{F}$ can be obtained by direct calculations, 
\begin{eqnarray}
\label{dFWcb}
\partial^j_{\mathsf{F}} W_{\rm cb}(\mathsf{F}_0) :=
\partial^j_{\mathsf{F}} W_{\rm cb}(\mathsf{F})\big|_{\mathsf{F} = \mathsf{F}_0} = - \sum_{\pmb\rho=(\rho_1, \cdots, \rho_j) \in (\L_{*})^j} V_{,\pmb\rho} ( \mathsf{F}_0\cdot\L_{*} ) \underbrace{\otimes{\rho}_1\otimes\cdots\otimes{\rho}_j}_{=:~\otimes\pmb\rho} ,
\end{eqnarray}
where $\otimes$ denotes the standard Kronecker product.

We then introduce the following accuracy measure of the virial stress: 
\begin{align}\label{training_V_full}
\vfit_j := \big|\partial^{j+1}_{\mathsf{F}} \Wcb(\mathsf{I}) - \partial^{j+1}_{\mathsf{F}} \Wcb^{\rm ML}(\mathsf{I}) \big| :=\Bigg| \sum_{\pmb\rho \in (\L_{*})^{j+1}} \big(V_{,\pmb\rho}(\mathsf{I}) - V^{\rm ML}_{,\pmb\rho}(\mathsf{I})\big) \otimes {\pmb\rho} \Bigg|,
\end{align}
for $j=1,\cdots,\mathfrak{n}-1$, where $\mathsf{I}$ is the identity matrix.

For the purpose of error analysis, we require the equilibrium of the single straight dislocation in infinite lattice. Following the results in \cite[Theorem 2.1]{chen19}, the corresponding energy-difference functional reads
\begin{eqnarray}\label{energy-difference}
\E^{\L}(y) := \sum_{\ell\in\Lambda} V_{\ell}\big(Dy(\ell)\big),
\end{eqnarray}
where $V_{\ell}({\bf g}) := V\big({\bf g}\big) - V\big(Dy^{\rm CLE}(\ell)\big)$.

The corresponding variational problem for the equilibrium state is
\begin{equation}\label{eq:variational-problem}
\bar{y}^{\L} \in \arg\min \big\{ \E^{\L}(y),~y-y^{\rm CLE} \in \UsH(\L) \big\},
\end{equation}
where ``$\arg\min$'' is understood as the set of local minimizers and the discrete energy space for infinite lattice
\begin{align}\label{space:UsH}
\UsH(\L) := \big\{u:\L\rightarrow\mathbb{R}^{3} ~\big\lvert~ \|Du\|_{\ell^2_{\mathcal{N}}(\L)}<\infty \big\}.
\end{align}

We will need a strong stability condition~\cite{chen19, Ehrlacher16} as well as qualitative information about the equilibrium, that is, 
\begin{eqnarray}\label{as:LS}
	\exists~\bar{c}>0~{\rm s.t.}~ 
		\big\< \delta^2\E^{\L}(\bar{y}^{\L}) v , v\big\> \geq \bar{c} \|Dv\|^2_{\ell^2} \qquad\forall~v\in\UsH(\L).
\end{eqnarray}

Before introducing the equilibrium of multiple dislocation configurations in $\L_N$, we define a family of defect core truncation operators $\{\Pi_{R}^{i}\}^{n_{\D}}_{i=1}$. Let $\eta \in C^1(\R^2;[0,1])$ be a cut-off function satisfying $\eta(x)=1$ for $|x|\leq 4/6$ and $\eta(x)=0$ for $|x|\geq 5/6$. We denote $\T_{\L}$ as the {\it canonical triangulation} of $\R^2$ whose nodes are the reference sites $\L$ (cf.~\cite[Section 2.1]{Ehrlacher16}). Let $Iu$ be the piecewise affine interpolant of $u$ with respect to $\T_{\L}$, and $A_{R}:=B_{5R/6}\setminus B_{4R/6}$ be an annulus, then we can define the truncation operator by
\begin{eqnarray}\label{eq:trun_op_def}
\Pi^{i}_{R}u(\ell):= \eta\Big(\frac{\ell - x^{\rm core}_i}{R}\Big)\big(u(\ell)-a^{i}_{R} \big), \quad \textrm{where} \quad a^{i}_{R}:=\bbint_{x^{\rm core}_i + A_R} Iu(x)\dx.
\end{eqnarray}
In particular, we denote $\Pi_{R}$ as the case that the core is placed at the origin.

\subsection{Proof of Assumption~\ref{ass:existence}}
\label{sec:apd:as}

As discussed in Remark~\ref{re:assump}, Assumption~\ref{ass:existence} is a conceptually straightforward but technically involved extension of our previous work~\cite[Theorem 2.1]{ortner2022framework} from point defects to dislocations. Our method of proof would lead to a constant $L_0$ that may depend on the number of defect cores $n_{\D}$, which is the most difficult gap to fill towards a rigorous proof of Assumption~\ref{ass:existence}.  In this section, we aim to present a sketch of the proof, focusing on the scenario where the number of cores is bounded.

We first define an approximated solution (predictor) to the variational problem~\eqref{eq:variational-problem-per} for dislocation configuration $\D$ with truncation radius $R=L_{\D}/3$ to be 
\begin{eqnarray}\label{eq:z}
z(\ell) = y^{\rm CLE}_{\rm per}(\ell) + \sum_{(x^{\rm core}_i, \bv_i) \in \D} \Pi_{R} \bar{u}^{\L}(\ell-x^{\rm core}_i; \bv_i), \qquad \forall \ell \in \L_N,
\end{eqnarray}
where $\Pi_{R}$ is defined by \eqref{eq:trun_op_def} with the core positioned at the origin. We then give an estimate on the residual of such an approximated solution in terms of $L_{\D}$. To be more precise, we want to prove that there exists a constant $L_0>0$ such that, for $L_{\D} > L_0$, 
\begin{eqnarray}\label{eq:res_estimate}
\big|\<\delta\E(z), v\>\big| \leq C \sqrt{n_{\D}} \cdot L_{\D}^{-1}\cdot \|D v\|_{\ell^2(\L_N)},
\end{eqnarray}
where the constant $C$ is independent of $N, n_{\D}, L_{\D}$.

As a matter of fact, for any $v \in \Us^{\per}_N$, we define
\begin{eqnarray}\label{eq:decomp_v_res}
v_i := \Pi^{i}_{r} v \quad \textrm{for}~i=1,\ldots,n_{\D}, \quad \textrm{and} \quad v_0:=v-\sum_{i=1}^{n_{\D}}v_i,
\end{eqnarray}
where $r:=R+1=L_{\D}/3+1$ and the defect core truncation operator $\Pi^{i}_{r}$ is defined by \eqref{eq:trun_op_def}.

We then decompose the residual into three parts
\begin{align}\label{eq:T12}
    \<\delta\E(z), v\> &= \sum_{i=0}^{n_{\D}}\<\delta\E(z), v_i\> \nonumber \\
    &= \<\delta \E(z), v_0\> + \sum^{n_{\D}}_{i=1} \<\delta\E(z)-\delta\E\big(y^{\rm CLE}_{\rm per} + T_N^{\rm per} \bar{u}^{\L}(\cdot - \ell_i)\big), v_i\> \nonumber \\
    &\hskip2.5cm + \sum^{n_{\D}}_{i=1} \<\delta\E\big(y^{\rm CLE}_{\rm per} + T_N^{\rm per} \bar{u}^{\L}(\cdot - \ell_i)\big), v_i\> \nonumber \\[1ex]
    &=: T_1 + T_2 + T_3,
\end{align}
where the operator $T_N^{\rm per}$ maps the displacements from $\Us^{1,2}(\L)$ to $\Us^{\rm per}_{N}(\L_N)$. The construction of $T_N^{\rm per}$ follows a similar procedure to that of training domains discussed in Section~\ref{sec:sub:gen}. 

Note that $r:=R+1=L_{\D}/3+1$. The term $T_1$ can be estimated by analyzing the residual of the linear elasticity predictor $u^{\rm CLE}_{\rm per}$. Following the analysis of~\cite[Lemma 4.3]{2014-dislift} and adapting it to the periodic setting, we can obtain that $|T_1|\leq Cr^{-2}\cdot\|Dv\|_{\ell^2(\L_N)}$. For the last two terms, following the proof of \cite[Lemma 6.3]{ortner2022framework}, we have $|T_2| \leq C\sqrt{n_{\D}} \cdot L_{\D}^{-1}\log(L_{\D}) \cdot\|Dv\|_{\ell^2(\L_N)}$ and $|T_3|\leq C\sqrt{n_{\D}} \cdot R^{-1}\log(R)\cdot\|Dv\|_{\ell^2(\L_N)}$. Hence, taking into account these estimates with \eqref{eq:T12}, we can obtain the following consistency
\[
\big|\<\delta\E(z), v\>\big| \leq C \sqrt{n_{\D}} \cdot L_{\D}^{-1}\log(L_{\D}) \cdot \|D v\|_{\ell^2(\L_N)}.
\]

To apply the inverse function theorem~\cite[Lemma A.1]{ortner2022framework}, we then proceed to prove that $\delta^2\E(z)$ is positive. This result employs the ideas similar to those used in the proofs of \cite[Theorem 7.7]{Ehrlacher16} and \cite[Lemma 5.2]{2014-dislift}, modified here to an periodic setting and extended to cover the case of multiple straight dislocations. 

We consider the following decomposition
\begin{align}\label{eq:splitS}
\<\delta^2\E(z) v, v\> &= \sum_{i,j=0}^{n_{\D}} \<\delta^2\E(z) v_i, v_j\> \nonumber \\
&= \<\delta^2\E(z) v_0, v_0\> + \sum_{i=1}^{n_{\D}}  \<\delta^2\E(z) v_i, v_i\> + 2\sum_{i=1}^{n_{\D}} \<\delta^2\E(z) v_0, v_i\> \nonumber \\[1ex]
&=: S_1 + S_2 + S_3,
\end{align}
where the first one is related to the stability of homogeneous lattice (phonon stability)~\cite[Proposition 6.1]{ortner2022framework}, the second one and the third one can be estimated by following the analysis in \cite{2014-dislift}
as well as the strongly stable of single dislocation shown in \eqref{as:LS}.

Hence, we can prove that, there exists a constant $L_0>0$ such that, for $L_{\D} > L_0$, there exists $\bar{c}_{0} \geq \bar{c}/2$ so that
\begin{eqnarray}\label{eq:exist:stab}
\<\delta^2\E(z)v, v\> \geq \bar{c}_{L_{\D}} \|Dv\|^2_{\ell^2(\L_{N})} \qquad \forall~v\in \Us^{\per}_N.  
\end{eqnarray}
We are aware of no argument that allows us to give a quantitative or even qualitative estimate on the magnitude of $L_0$. 

Applying the inverse function theorem~\cite[Lemma A.1]{ortner2022framework}, we can obtain that there exist $L_0>0$, where $\D$ satisfies $L_{\D}\geq L_0$, and $z$ is an approximated solution defined by \eqref{eq:z} corresponding to $\D$. It follows that for any $v \in \Us_N^{\per}$, there exists $\omega \in \Us_N^{\per}$ with 
\begin{eqnarray}\label{eq:Dw}
\|D\omega\|_{\ell^2(\L_N)} \leq C \sqrt{n_{\D}} \cdot L_{\D}^{-1} \log(L_{\D}),
\end{eqnarray}
such that
\begin{align*}
\<\delta\E(z+\omega), v\>=0, \quad \<\delta^2\E(z+\omega)v, v\> \geq \frac{\bar{c}_{L_{\D}}}{2}\|Dv\|^2_{\ell^2}.
\end{align*}
Writing $\bar{y}:=z+\omega$ yields the stated result of the Assumption~\ref{ass:existence}. 

\subsection{Proof of the generalisation analysis}
\label{sec:apd:proof}

We are ready to give the detailed proof of the generalisation analysis (Theorem~\ref{them:geometry}), which is the main result in this paper.

\begin{proof}
Applying the framework of the {\it a priori} error estimates in \cite{chen17, co2011, colz2012, colz2016}, we divide the proof into several steps in order to apply the inverse function theorem~\cite[Lemma A.1]{ortner2022framework}.

{\it 1. Stability:} For any $v \in \Us^{\per}_N$, we consider the stability of
\begin{align}\label{eq:thm2:splitS}
\<\delta^2 \E^{\rm ML}(\bar{y})v, v\> &=  \<\delta^2 \E(\bar{y})v, v \> + \big( \<\delta^2 \E^{\rm ML}(\bar{y})v, v \> - \<\delta^2 \E(\bar{y})v, v \> \big) \nonumber \\[1ex]
&=: S_1 + S_2.
\end{align}
From the Assumption~\ref{ass:existence}, we can obtain that 
\[
S_1:=\<\delta^2\E(\bar{y})v, v\> \geq c_0\|Dv\|^2_{\ell^2}.
\]

We split the test function $v$. For $L \leq L_{\D}$, let $r:=L/3+1$, we define
\begin{eqnarray}\label{eq:decomp_v}
v_i := \Pi^{i}_{r} v \quad \textrm{for}~i=1,\ldots,n_{\D}, \quad \textrm{and} \quad v_0:=v-\sum_{i=1}^{n_{\D}}v_i.
\end{eqnarray}
It is shown in \cite[Lemma A.2]{ortner2022framework} that $\|D v_i\|_{\ell^2(\L_{N})} \leq C\|D v\|_{\ell^2(\L_{N})}$ for $i=1,\ldots,n_{\D}$. 

The term $S_2$ can be further split into three parts
\begin{align}\label{eq:thm2:splitS2}
\big\<\big(\delta^2{\E}^{\rm ML}(\bar{y})-\delta^2\E(\bar{y})\big) v, v\big\> =&~ \sum_{i,j=0}^{n_{\D}} \big\<\big(\delta^2{\E}^{\rm ML}(\bar{y})-\delta^2\E(\bar{y})\big) v_i, v_j \big\> \nonumber \\[1ex]
=& ~\big\<\big(\delta^2{\E}^{\rm ML}(\bar{y})-\delta^2\E(\bar{y})\big) v_0, v_0\big\> + \sum_{i=1}^{n_{\D}}  \big\<\big(\delta^2{\E}^{\rm ML}(\bar{y})-\delta^2\E(\bar{y})\big) v_i, v_i \big\> \nonumber \\ 
&+ 2\sum_{i=1}^{n_{\D}} \big\<\big(\delta^2{\E}^{\rm ML}(\bar{y})-\delta^2\E(\bar{y})\big) v_0, v_i \big\> \nonumber \\[1ex]
=:& ~S_{21} + S_{22} + S_{23},
\end{align}
where we have ensured that supp$(v_i)$ for $i=1,\ldots,n_{\D}$ only overlaps with supp$(v_0)$ by the choice of $r$ and therefore all other cross-terms vanish.

To simply the notation, we denote $x_0(\ell):=\ell$, $\tilde{y}:=y^{\rm CLE}_{\rm per}+\omega$ and $\tilde{u}:=u^{\rm CLE}_{\rm per}+\omega$. For the term $S_{21}$, for $L$ sufficiently large and $t\in[0,1]$, we can Taylor expand both $\delta^2\E$ and $\delta^2{\E}^{\rm ML}$ at the reference configuration
\begin{align}
&\big\<\big(\delta^2{\E}^{\rm ML}(\tilde{y})-\delta^2\E(\tilde{y})\big)v_0, v_0\big\> \nonumber \\[1ex]
=~& \big\<\big(\delta^2 {\E}^{\rm ML}(x_0)-\delta^2 \E(x_0)\big)v_0, v_0\big\> + \big\<\big(\delta^3 {\E}^{\rm ML}(x_0+t\tilde{u})-\delta^3 \E(x_0+t\tilde{u})\big)\tilde{u} v_0, v_0\big\> \nonumber \\[1ex]
=:&~S^{\rm (a)}_{21} + S^{\rm (b)}_{21}. 
\end{align}

To estimate the first term $S^{\rm (a)}_{21}$, for $L$ sufficiently large, it is easy to see that
\begin{align}\label{eq:s21a}
|S^{\rm (a)}_{21}| 
\leq C \varepsilon^{\rm H} \cdot \|D  v_0\|^2_{\ell^2(\L_N)} \leq C \varepsilon^{\rm H} \cdot \|D v\|^2_{\ell^2(\L_N)}, 
\end{align}
where $\varepsilon^{\rm H}$ is defined by~\eqref{eq:FCD} and the last inequality follows from the fact that $\|D v_0\|_{\ell^2(\L_{N})} \leq C\|D v\|_{\ell^2(\L_{N})}$ shown in \cite[Lemma A.2]{ortner2022framework}.

For the term $S^{\rm (b)}_{21}$, from the definition given by \eqref{energy-difference-per}, similarly we can obtain
\begin{align}
|S^{\rm (b)}_{21}| &\leq C \sum_{\ell\in\L_N}\big|Du^{\rm CLE}_{\rm per}(\ell)+D\omega(\ell)\big|_{\mathcal{N}}\cdot\big|Dv_0(\ell)\big|^2_{\mathcal{N}} \nonumber \\[1ex]
&\leq C \big(\|Du^{\rm CLE}_{\rm per}\|_{\ell^{\infty}(\textrm{supp}(v_0))} + \|D\omega\|_{\ell^{\infty}(\textrm{supp}(v_0))}\big) \cdot \|D v_0\|^2_{\ell^2(\textrm{supp}(v_0))} \nonumber \\[1ex]
&\leq C L^{-1} \cdot \|D v\|^2_{\ell^2(\L_N)},
\end{align}
where the last inequality follows from the estimates \eqref{eq:Dw} and \eqref{eq:decay_u0}.

Let $\Pi_R\bar{y}^{\L}:= y^{\rm CLE}_{\rm per} + \Pi_R \bar{u}^{\L}$. To estimate $S_{22}$, recall the definition of the predictor by \eqref{eq:z} and the construction of $v_i$, for each $i=1,\ldots,n_{\D}$ and $L$ sufficiently large, we have
\begin{align}
&\big\<\big(\delta^2{\E}^{\rm ML}(\bar{y})-\delta^2\E(\bar{y})\big) v_i, v_i\big\> \nonumber \\[1ex]
=&~ \big\<\big(\delta^2{\E}^{\rm ML}(\Pi_{R}\bar{y}^{\L}(\cdot-x^{\rm core}_i) )-\delta^2\E(\Pi_{R}\bar{y}^{\L}(\cdot-x^{\rm core}_i))\big) v_i, v_i\big\> \nonumber \\[1ex]
=&~ \big\<\big(\delta^2{\E}^{\rm ML}(\Pi_{R}\bar{y}^{\L}(\cdot-x^{\rm core}_i) )-\delta^2{\E}^{\rm ML}(\bar{y}_L(\cdot-x^{\rm core}_i))\big) v_i, v_i\big\> \nonumber \\
&+ \big\<\big(\delta^2{\E}^{\rm ML}(\bar{y}_L(\cdot-x^{\rm core}_i) )-\delta^2\E(\bar{y}_L(\cdot-x^{\rm core}_i))\big) v_i, v_i\big\> \nonumber \\
&+ \big\<\big(\delta^2\E(\bar{y}_L(\cdot-x^{\rm core}_i) )-\delta^2\E(\Pi_{R}\bar{y}^{\L}(\cdot-x^{\rm core}_i))\big) v_i, v_i\big\> \nonumber \\[1ex]
\leq&~ C (\varepsilon^{\rm H} + L^{-1} + \|D \bar{y}^{\L} - D\bar{y}_L\|_{\ell^2(\L_L)}) \cdot \|D v_i\|^2_{\ell^2(\textrm{supp}(v_i))} \nonumber \\[1ex]
\leq&~ C (\varepsilon^{\rm H} + L^{-1}) \cdot \|D v\|^2_{\ell^2(\L_N)}, 
\end{align}
where the last inequality follows from the results in~\cite{Ehrlacher16}.

Noting that $\<\delta^2\E(y) v_0, v_i\> = \<\delta^2\E(y) (v-v_i), v_i\>$, the term $S_{23}$ can be estimated similarly by
\begin{align}\label{eq:S23}
    \big\<\big(\delta^2{\E}^{\rm ML}(\bar{y})-\delta^2\E(\bar{y})\big) v_0, v_i\big\>
    \leq C (\varepsilon^{\rm H}+L^{-1}) \cdot \|D v\|^2_{\ell^2(\L_N)}.
\end{align}

Hence, combining the estimates from \eqref{eq:thm2:splitS} to \eqref{eq:S23}, for $L$ sufficiently large and the matching condition $\varepsilon^{\rm H}$ sufficiently small, we have
\begin{eqnarray}\label{eq:thm2:stab}
\<\delta^2{\E}^{\rm ML}(\bar{y})v, v\> \geq \frac{c_0}{2}\|Dv\|^2_{\ell^2(\L_N)}.
\end{eqnarray}

{\it 2. Consistency:}
We estimate the consistency error, for any $v \in \Us_{N}^{\rm per}$, by
\begin{align}\label{eq:splitcons}
    \<\delta {\E}^{\rm ML}(\bar{y}), v\> &= \<\delta {\E}^{\rm ML}(\bar{y})-\delta\E(\bar{y}), v\> \nonumber \\[1ex]
    &= \sum^{n_{\D}}_{i=1} \<\delta {\E}^{\rm ML}(\bar{y})-\delta\E(\bar{y}), v_i\> + \<\delta {\E}^{\rm ML}(\bar{y})-\delta\E(\bar{y}), v_0\> \nonumber \\[1ex]
    &=: T_1 + T_2,
\end{align}
where $v_i, i=0,\ldots,n_{\D}$, are constructed by \eqref{eq:decomp_v}.

Let $\Pi_R\bar{y}^{\L}:= y^{\rm CLE}_{\rm per} + \Pi_R \bar{u}^{\L}$. To estimate $T_1$, for each $i=1, \ldots, n_{\D}$, we denote
\begin{align}\label{eq:cons1}
    T^{\rm (i)}_{1} :=& \<\delta {\E}^{\rm ML}(\bar{y})-\delta\E(\bar{y}), v_i\> \nonumber \\[1ex]
    =& \big\<\delta{\E}^{\rm ML}\big(\Pi_{R}\bar{y}^{\L}(\cdot-x^{\rm core}_i)\big)-\delta\E\big(\Pi_{R}\bar{y}^{\L}(\cdot-x^{\rm core}_i)\big), v_i\big\>. 
\end{align}
Given $\delta>0$, for $L$ sufficiently large, we have $\Pi_{R}\bar{y}^{\L} \in B_{\delta}(\bar{y}_L)$. Recalling the definition of $\varepsilon^{\rm F}$ by \eqref{eq:ffit}, we can obtain 
\begin{eqnarray}\label{eq:T1a}
|T_1| \leq \sum^{n_{\D}}_{i=1} |T^{\rm (i)}_1| \leq  C \sqrt{n_{\D}} \cdot \varepsilon^{\rm F} \cdot \|D v\|_{\ell^2(\L_N)}.
\end{eqnarray}

In the next steps of our analysis, we draw from techniques employed in the study of the Cauchy--Born continuum model~\cite{co2013, wang2021priori} and blended atomistic-to-continuum methods~\cite{mlco2013, li2016analysis, wang2023efficient, liao2021adaptive, wang2021posteriori}. For $v \in \Us^{\rm per}_{N}$, we introduce two smooth interpolant operators. The first operator $\mathcal{I}: \Us^{\rm per}_N \rightarrow C^{2,1}$, is employed to establish the regularity of $v$~\cite[Section 2.2.1]{li2016analysis} while the second one $\mathcal{J}: \Us_{N}^{\rm per} \rightarrow \dot{H}^1_{\rm per}$, is utilized for the construction of the so-called {\it smeared bond integrals}~\cite[Section 5.1]{li2016analysis}. The dual operator of $\mathcal{J}$ is then represented as $\mathcal{J}^*: \dot{H}^{-1}_{\rm per} \rightarrow (\Us_N^{\rm per})^*$. It is shown in~\cite[Lemma 5.1]{li2016analysis} that $\|\nabla (\mathcal{J}v)\|_{L^2} \leq C\|Dv\|_{\ell^2}$.

Recalling the Cauchy-Born elastic energy density functional defined by \eqref{eq:cbW}, we denote the corresponding Cauchy-Born energy as $\E_{\rm cb}$ and $\E^{\rm ML}_{\rm cb}$. To further simplify the notation, we denote $\mathcal{F}:=\delta\E$, and similarly we define $\mathcal{F}_{\rm cb}$, $\mathcal{F}^{\rm ML}$, and $\mathcal{F}^{\rm ML}_{\rm cb}$. Note that $\mathcal{F}_{\rm cb}, \mathcal{F}_{\rm cb}^{\rm ML} \in \dot{H}^{-1}_{\rm per}$. 
The functionals $\mathcal{F}_{\rm cb}$ (and analogously $\mathcal{F}_{\rm cb}^{\rm ML}$) are {\it defined} via the identity: 
\[
\big\< \mathcal{J}^{*} \mathcal{F}_{\rm cb}(y), v \big\> = \big\< \mathcal{F}_{\rm cb}(y), \mathcal{J}v \big\> = \int_{\Omega_N} \partial_{\rm F} W_{\rm cb}\big(\nabla (\mathcal{I}y) \big) : \nabla (\mathcal{J}v) \dx, 
\qquad \forall v\in\Us_N^{\rm per}.
    \]

We first split $T_2$ into three parts:
\begin{align}\label{eq:split_T2}
    T_2 =&~\big\< \mathcal{F}(y^{\rm CLE}_{\rm per}+\omega) -  \mathcal{F}^{\rm ML}(y^{\rm CLE}_{\rm per}+\omega), v_0 \big\> \nonumber \\[1ex]
    =&~\big\< \mathcal{F}(y^{\rm CLE}_{\rm per}+\omega) -  \mathcal{J}^{*}\mathcal{F}_{\rm cb}(y^{\rm CLE}_{\rm per}+\omega), v_0 \big\> \nonumber \\
    &+ \big\< \mathcal{J}^{*} \mathcal{F}^{\rm ML}_{\rm cb}(y^{\rm CLE}_{\rm per}+\omega) -  \mathcal{F}^{\rm ML}(y^{\rm CLE}_{\rm per}+\omega), v_0 \big\> \nonumber \\
    &+ \big\< \mathcal{J}^{*} \mathcal{F}_{\rm cb}(y^{\rm CLE}_{\rm per}+\omega) -  \mathcal{J}^{*} \mathcal{F}^{\rm ML}_{\rm cb}(y^{\rm CLE}_{\rm per}+\omega), v_0 \big\> \nonumber \\[1ex]
    =:&~T_{21} + T_{22} + T_{23}.
\end{align}

Let $\tilde{y} := y^{\rm CLE}_{\rm per}+\omega$ and $\tilde{u} := u^{\rm CLE}_{\rm per}+\omega$. The first two terms, $T_{21}$ and $T_{22}$, can be estimated by the Cauchy-Born (continuum) modeling error~\cite[Lemma 4.5]{co2013}:
\begin{align}\label{eq:T21}
|T_{21}| + |T_{22}| &=  
\Big| \big\< \mathcal{F}(\tilde{y}) - \mathcal{J}^{*} \mathcal{F}_{\rm cb}(\tilde{y}), v_0 \big\> \Big|  +  \Big| \big\< \mathcal{J}^{*} \mathcal{F}^{\rm ML}_{\rm cb}(\tilde{y}) -  \mathcal{F}^{\rm ML}(\tilde{y}), v_0 \big\> \Big| \nonumber \\[1ex]
&\leq C \big(\|\nabla^3 (\mathcal{I}\tilde{y})\|_{L^2} + \|\nabla^2 (\mathcal{I}\tilde{y})\|^2_{L^4} \big) \cdot \|\nabla (\mathcal{J}v_0)\|_{L^2} \nonumber \\[1ex]
&\leq C L^{-2} \cdot \|Dv_0\|_{\ell^2} \leq C L^{-2} \cdot \|Dv\|_{\ell^2},
\end{align}
where the last line follows from the fact that $|\nabla (\mathcal{I}y^{\rm CLE}_{\rm per})|\leq C|x|^{-1}$, the estimate $\|\nabla (\mathcal{J}v_0)\|_{L^2} \leq C\|Dv_0\|_{\ell^2}$ by~\cite[Lemma 5.1]{li2016analysis} and $\|D v_0\|_{\ell^2} \leq C\|D v\|_{\ell^2}$ shown in \cite[Lemma A.2]{ortner2022framework}. 

To estimate the last term $T_{23}$, we have 
\begin{align}\label{eq:T23_1}
&~\big\< \mathcal{J}^{*} \mathcal{F}^{\rm ML}_{\rm cb}(\tilde{y}) -  \mathcal{J}^{*} \mathcal{F}_{\rm cb}(\tilde{y}), v_0 \big\> \nonumber \\[1ex]
=&~\big\< \mathcal{F}^{\rm ML}_{\rm cb}(\tilde{y}) - \mathcal{F}_{\rm cb}(\tilde{y}), \mathcal{J}v_0 \big\> \nonumber \\[1ex]
=&\int \Big( \partial_{\rm F} W^{\rm ML}_{\rm cb}\big(\nabla (\mathcal{I}\tilde{y})\big) - \partial_{\rm F} W_{\rm cb}\big(\nabla (\mathcal{I}\tilde{y}) \big) \Big) : \nabla (\mathcal{J}v_0) \dx.
\end{align}
Note that $D\bar{u}$ and hence also $\nabla\tilde{u}$ are small and smooth in supp$(v_0)$ for $L$ sufficiently large. Hence, we can Taylor expand $\partial_{\mathsf{F}}{W}^{\rm ML}_{\rm cb}$ and $\partial_{\mathsf{F}}{W}_{\rm cb}$ at the reference 
\begin{align}\label{eq:T23_2}
    &\partial_{\rm F} W^{\rm ML}_{\rm cb}\big(\nabla (\mathcal{I}\tilde{y})\big) - \partial_{\rm F} W_{\rm cb}\big(\nabla (\mathcal{I}\tilde{y}) \big) \nonumber \\[1ex]
    =~& \big(\partial^2_{\rm F} W^{\rm ML}_{\rm cb}(\mathsf{I}) - \partial^2_{\rm F} W_{\rm cb}(\mathsf{I}) \big): \nabla (\mathcal{I}\tilde{u}) \nonumber \\ 
    &+ \frac{1}{2}\big(\partial^3_{\rm F} W^{\rm ML}_{\rm cb}(\mathsf{I}) - \partial^3_{\rm F} W_{\rm cb}(\mathsf{I}) \big): \big(\nabla (\mathcal{I}\tilde{u})\big)^{\otimes 2} \nonumber \\
    &+ \frac{1}{6}\big(\partial^4_{\rm F} W^{\rm ML}_{\rm cb}(\mathsf{I}+t\nabla(\mathcal{I}\tilde{u}) ) - \partial^4_{\rm F} W_{\rm cb}(\mathsf{I}+t\nabla(\mathcal{I}\tilde{u}) ) \big): \big(\nabla (\mathcal{I}\tilde{u})\big)^{\otimes 3},
\end{align}
where $t \in [0,1]$ and $v^{\otimes k} := v \otimes \ldots \otimes v$ ($k$ times).

Hence, taking \eqref{eq:T23_2} into account with \eqref{eq:T23_1}, we have
\begin{align}\label{eq:T23}
    |T_{23}| &\leq C \big( \vfit_1 \|\nabla (\mathcal{I}\tilde{u})\|_{L^2} + \vfit_2 \|\nabla (\mathcal{I}\tilde{u})\|^2_{L^4} + \|\nabla (\mathcal{I}\tilde{u})\|^3_{L^6} \big) \cdot \|\nabla (\mathcal{J}v_0)\|_{L^2} \nonumber \\[1ex]
    &\leq C \big(\vfit_1 \log^{1/2}(L_{\D}) + \vfit_2 L^{-1} + L^{-2}\big) \cdot \|Dv\|_{\ell^2},
\end{align}
where the last inequality follows from the fact that $|\nabla (\mathcal{I}\tilde{u})|\leq C|x|^{-1}$, the estimate $\|\nabla (\mathcal{J}v_0)\|_{L^2} \leq C\|Dv_0\|_{\ell^2}$ by~\cite[Lemma 5.1]{li2016analysis} and $\|D v_0\|_{\ell^2} \leq C\|D v\|_{\ell^2}$ shown in \cite[Lemma A.2]{ortner2022framework}. 

From the above analysis, it is easy to see that the higher-order ($j\geq 3$) derivatives of the virial does not inherently lead to a systematic improvement in convergence rates with respect to $L$, as the Cauchy-Born (continuum) modeling error ($L^{-2}$) is of the same order as the first term that we neglected in the expansion of the virial.

In summary, combining from \eqref{eq:splitcons} to \eqref{eq:T23}, we can obtain 
\begin{eqnarray}\label{eq:thm2:cons}
\<\delta {\E}^{\rm ML}(\bar{y}), v\> \leq C \sqrt{n_{\D}} \cdot \big( \varepsilon^{\rm F} + \log^{1/2}(L_{\D})\cdot\vfit_1 + L^{-1}\cdot \varepsilon^{\rm V}_{2} + L^{-2}\big)\cdot \|D v\|_{\ell^2(\L_N)}.
\end{eqnarray}

{\it 3. Application of inverse function theorem:} Applying the framework of the {\it a priori} error estimates in \cite{ortner2022framework, 2021-qmmm3, olson2019theoretical}, with the stability \eqref{eq:thm2:stab} and consistency \eqref{eq:thm2:cons}, we can apply the inverse function theorem to obtain, for $L$ sufficiently large and the matching conditions $\ffit, \vfit_1, \varepsilon^{\rm H}$ sufficiently small, the existence of a solution $\bar{u}^{\rm ML}$ to \eqref{eq:variational-problem-per}, and the estimate
\[
\|D\bar{y} - D\bar{y}^{\rm ML}\|_{\ell^2(\L_N)} \leq C^{\rm G} \sqrt{n_{\D}} \cdot \big( \varepsilon^{\rm F} + \log^{1/2}(L_{\D})\cdot\vfit_1 + L^{-1}\cdot \varepsilon^{\rm V}_{2} + L^{-2} \big),
\]
where $C^{\rm G}$ is independent of $N, n_{\D}, L_{\D}$ and $L$. This completes the proof of \eqref{eq:geoerr}.


{\it 4: Error in energy:} Next, we estimate the error in the energy~\cite{ortner2022framework, 2021-qmmm3}. Recall the definition of the predictor $z$ by \eqref{eq:z}, we first spilt the error in energy into two parts
\begin{eqnarray}\label{eq:splitE}
\big| \E(\bar{y}) - {\E}^{\rm ML}(\bar{y}^{\rm ML}) \big| \leq \big| \E(\bar{y}) - \E(z)\big| + \big| \E(z) - {\E}^{\rm ML}(\bar{y}^{\rm ML}) \big| =: E_1 + E_2.
\end{eqnarray}
The term $E_1$ can be bounded by
\begin{align}\label{eq:E1}
\big|\E(\bar{y}) - \E(z)\big| &= \Big| \int_0^1 \big\<\delta\E\big((1-s)\bar{y}+sz\big), \bar{y}-z \big\> \ds \Big| \nonumber \\[1ex]
&= \Big| \int_0^1 \big\<\delta\E\big((1-s)\bar{y}+sz\big)-\delta\E(\bar{y}), \bar{y}-z \big\> \ds \Big| \nonumber \\[1ex]
&\leq C M_1 \cdot \|D \bar{y} - Dz \|^2_{\ell^2(\L_N)} \leq C n_{\D} \cdot L^{-2},
\end{align}
where $M_1$ is the uniform Lipschitz constant of $\delta\E$.

To estimate $E_2$, by applying the technique used in \eqref{eq:E1}, similarly we can obtain  
\begin{align}\label{eq:E2}
\big| {\E}^{\rm ML}(\bar{y}^{\rm ML}) - \E(z)\big| &\leq \big| {\E}^{\rm ML}(\bar{y}^{\rm ML}) - {\E}^{\rm ML}(z)\big| + \big| {\E}^{\rm ML}(z) - \E(z)\big| \nonumber \\[1ex]
&\leq C n_{\D} \cdot \big( \|D\bar{y}^{\rm ML} - D\bar{y}\|^2_{\ell^2(\L_N)} + \|D\bar{y} - Dz\|^2_{\ell^2(\L_N)} + \varepsilon^{\rm E} \big) \nonumber \\[1ex]
&\leq C n_{D} \cdot \big(\|D\bar{y}^{\rm ML} - D\bar{y}\|^2_{\ell^2(\L_N)} + L^{-2} + \varepsilon^{\rm E} \big).
\end{align}

Combining \eqref{eq:splitE}, \eqref{eq:E1} and \eqref{eq:E2}, we obtain
\[
\big| \E(\bar{y}) - {\E}^{\rm ML}(\bar{y}^{\rm ML}) \big| \leq C^{\rm E} n_{\D} \cdot \Big( \big( \varepsilon^{\rm F} + \log^{1/2}(L_{\D})\cdot\vfit_1 + L^{-1}\cdot \varepsilon^{\rm V}_{2}\big)^2 + L^{-2} + \varepsilon^{\rm E} \Big),
\]
which completes the proof of Theorem \ref{them:geometry}.
\end{proof}

\section{Predictors}
\label{sec:apd:u0}

In this part, we briefly recall the predictor of a single dislocation on infinite lattice $\L$, which requires a small modification to the standard CLE solution. Let $\hat{x} \in \mathbb{R}^2$ be the position of dislocation core and $\Gamma := \{x \in \mathbb{R}^2~|~x_2 = \hat{x_2}, x_1 \geq \hat{x_1}\}$ be the ``branch-cut", with $\hat{x}$ chosen such that $\Gamma \cap \Lambda = \emptyset$. We define the far-field predictor $u_0$ by
\begin{eqnarray}\label{predictor-u_0-dislocation}
u_0(x):=\ulin(\xi^{-1}(x)),
\end{eqnarray}
where $\ulin \in C^\infty(\R^2 \setminus \Gamma; \R^d)$ is the solution of continuum linear elasticity (CLE)
\begin{align}\label{CLE}
\nonumber
\mathbb{C}^{j\beta}_{i\alpha}\frac{\partial^2 u^{\rm lin}_i}{\partial x_{\alpha}\partial x_{\beta}} &= 0 \qquad \text{in} ~~ \R^2\setminus \Gamma,
\\
u^{\rm lin}(x+) - u^{\rm lin}(x-) &= -\burg \qquad \text{for} ~~  x\in \Gamma \setminus \{\hat{x}\},
\\
\nonumber
\nabla_{e_2}u^{\rm lin}(x+) - \nabla_{e_2}u^{\rm lin}(x-) &= 0 \qquad \text{for} ~~  x\in \Gamma \setminus \{\hat{x}\},
\end{align}
where the forth-order tensor $\mathbb{C}$ is the linearised Cauchy-Born tensor (derived from the potential $\Vhom$, see \cite[\S~7]{Ehrlacher16} for more detail),
\begin{eqnarray}
\xi(x)=x-\burg_{12}\frac{1}{2\pi}
\eta\left(\frac{|x-\hat{x}|}{\hat{r}}\right)
\arg(x-\hat{x}),
\end{eqnarray}
with $\arg(x)$ denoting the angle in $(0,2\pi)$ between $x$ and
$\burg_{12} = (\burg_1, \burg_2) = (\burg_1, 0)$, and
$\eta\in C^{\infty}(\R)$ with $\eta=0$ in $(-\infty,0]$ and $\eta=1$ in
$[1,\infty)$ which removes the singularity. It is widely recognized that the gradient of the displacement field $u_0$ follows $r^{-1}$ with respect to the distance from $\hat{x}$.

In order to model dislocations, the homogeneous site potential $V$ must be invariant under lattice slip. Following \cite{Ehrlacher16}, we define the slip operator $S_0$ acting on the displacements $w: \Lambda \rightarrow \mathbb{R}^m$, by ($\mathsf{b}_{12}$ represents the projection of the Burger's vector to the $(x_1,x_2)$ plane)
\[
S_0w(x) := \begin{cases}
~~w(x) \qquad \qquad \qquad &x_2 > \hat{x}_2 \\
~~w(x-\mathsf{b}_{12}) - \mathsf{b} \qquad &x_2 < \hat{x}_2
\end{cases}.
\]

We may then formulate the slip invariance condition by defining a mapping $S$, where $S$ is an $\ell^2$-orthogonal operator with dual $S^{*}=S^{-1}$ by
\[
Su(\ell) := \begin{cases} ~~u(\ell) \qquad &\ell_2 > \hat{x}_2 \\
~~u(\ell-\mathsf{b}_{12}) \qquad &\ell_2 < \hat{x}_2
\end{cases}, \qquad 
S^{*}u(\ell) = \begin{cases} ~~u(\ell) \qquad &\ell_2 > \hat{x}_2 \\
~~u(\ell+\mathsf{b}_{12}) \qquad &\ell_2 < \hat{x}_2
\end{cases}.
\]
The slip invariance condition can now be expressed  as
\begin{equation}\label{slip}
V\big(D(u_0 + u)(\ell)\big) = V\big(S^{*}DS_0(u_0 + u)(\ell)\big), \qquad \forall~\ell \in \Lambda, u \in \UsH(\L),
\end{equation}
where $u_0$ is defined by \eqref{predictor-u_0-dislocation}.

In our analysis we require that applying the slip operator to the predictor map $u_0$ yields a smooth function in the half-space $\Omega_\Gamma = \{x_1 \geq \hat{x}_1 + \hat{r}+\mathsf{b}_{1}\}$.
It is therefore natural to define (likewise to \cite{Ehrlacher16}) the elastic strains
\begin{equation}\label{strain}
e(\ell) := \big(e_{\rho}(\ell)\big)_{\rho \in \L-\ell}, \qquad e_{\rho}(\ell) = \begin{cases} ~~S^{*}D_{\rho}S_0u_0(\ell) \qquad &\ell \in \Omega_{\Gamma} \\
~~D_{\rho}u_0(\ell) \qquad  &\ell \notin \Omega_\Gamma
\end{cases},
\end{equation}
and the analogous definition for corrector $u$
\begin{equation}\label{fancy_D}
Du(\ell) := \big(D_{\rho}u(\ell)\big)_{\rho \in \L-\ell}, \qquad D_{\rho}u(\ell) = \begin{cases} ~~S^{*}D_{\rho}Su(\ell) \qquad &\ell \in \Omega_{\Gamma} \\
~~D_{\rho}u(\ell) \qquad &\ell \notin \Omega_\Gamma
\end{cases}.
\end{equation}
Using this notation, the slip invariance condition \eqref{slip} may be written as, for $u\in\UsH(\L)$,
\begin{equation}\label{slip_strain}
V\big(D(u_0 + u)(\ell)\big) = V\big(e(\ell) + Du(\ell)\big).
\end{equation}

The following lemma, proven in \cite{chen19}, is a straightforward extension of \cite[Lemma 3.1]{Ehrlacher16}.

\begin{lemma}
If the predictor $u_0$ is defined by \eqref{predictor-u_0-dislocation} and $e(\ell)$ is given by \eqref{strain}, then there exists a constant $C$ such that 
\begin{equation}\label{eq:decay_u0}
    |e_{\sigma}(\ell)| \leq C |\sigma|\cdot |\ell|^{-1} \qquad {\rm and} \qquad |D_{\rho} e_{\sigma}(\ell)| \leq C |\rho| \cdot |\sigma|\cdot |\ell|^{-2}.
\end{equation}
\end{lemma}

\section{The Atomic Cluster Expansion}
\label{sec:ACE}
We briefly review the construction of the ACE potential, but refer to \cite{2019-ship1,Drautz19,lysogorskiy2021performant,witt2023acepotentials} for further details. 
Given a {\it correlation order} $\mathcal{N} \in \N$, we first write the ACE site potential in the form of an {\it atomic body-order expansion}, $\displaystyle V^{\rm ACE} \big(\{\pmb{g_j}\}\big) = \sum_{N=0}^{\mathcal{N}} \frac{1}{N!}\sum_{j_1 \neq \cdots \neq j_N} V_N(\pmb{g}_{j_1}, \cdots, \pmb{g}_{j_N})$, where the $N$-body potential $V_N : \R^{dN} \rightarrow \R$ can be approximated by using a tensor product basis \cite[Proposition 1]{2019-ship1},
\begin{align*}
\phi_{\pmb{n\ell m}}\big(\{\pmb{g}_j\}_{j=1}^N\big) :=\prod_{j=1}^N\phi_{n_j\ell_j m_j}(\pmb{g}_j) 
\quad & {\rm with}\quad 
\phi_{n\ell m}(\pmb{r}):=P_n(r)Y^{m}_{\ell}(\hat{r}) , ~~\pmb{r}\in\R^d,~r=|\pmb{r}|,~\hat{r}=\pmb{r}/r ,
\end{align*}
where $P_n,~n=0,1,2,\cdots$ are radial basis functions (for example, Jacobi polynomials), and $Y_{\ell}^m,~\ell = 0,1,2,\cdots,~m=-\ell,\cdots,\ell$ are the complex spherical harmonics.
The basis functions are further symmetrised to a permutation invariant form,
\begin{eqnarray*}
\sum_{(\pmb{n,\ell,m})~{\rm ordered}} \sum_{\sigma\in S_N} \phi_{\pmb{n\ell m}} \circ \sigma ,
\end{eqnarray*}  
where $S_N$ is the collection of all permutations, and by $\sum_{(\pmb{n,\ell,m})~{\rm ordered}}$ we mean that the sum is over all lexicographically ordered tuples $\big( (n_j, \ell_j, m_j) \big)_{j=1}^N$.
The next step is to incorporate the invariance under point reflections and rotations
\begin{eqnarray*}
\mathcal{B}_{\pmb{n\ell} i} = \sum_{\pmb{m}\in\mathcal{M}_{\pmb{\ell}}}\mathcal{U}_{\pmb{m}i}^{\pmb{n\ell}} \sum_{\sigma\in S_N} \phi_{\pmb{n\ell m}} \circ \sigma 
\quad{\rm with}\quad \mathcal{M}_{\pmb{\ell}} = \big\{\pmb{\mu}\in\Z^N ~|~ -\ell_{\alpha}\leq\mu_{\alpha}\leq\ell_{\alpha} \big\} ,
\end{eqnarray*}
where the coefficients $\mathcal{U}_{\pmb{m}i}^{\pmb{nl}}$ are given in \cite[Lemma 2 and Eq. (3.12)]{2019-ship1}.
It was shown in \cite{2019-ship1} that the basis defined above is explicit but computational inefficient. The so-called ``density trick" technique used in \cite{Bart10, Drautz19, Shapeev16} can transform this basis into one that is computational efficient. The alternative basis is 
\begin{eqnarray*}
B_{\pmb{n\ell} i} = \sum_{\pmb{m}\in\mathcal{M}_{\pmb{\ell}}}\mathcal{U}_{\pmb{m}i}^{\pmb{n\ell}} A_{\pmb{nlm}}
\quad{\rm with~the~correlations}\quad A_{\pmb{nlm}}:= \prod_{\alpha=1}^{N} \sum_{j=1}^{J} \phi_{n_\alpha l_\alpha m_{\alpha}}(\pmb{g}_j),
\end{eqnarray*}
which avoids both the $N!$ cost for symmetrising the basis as well as the $C_J^N$ cost of summation over all order $N$ clusters within an atomic neighbourhood.
The resulting basis set is then defined by
\begin{eqnarray}
\pmb{B}_N := \big\{ B_{\pmb{n\ell} i}  ~|~ (\pmb{n},\pmb{\ell})\in\N^{2N}~{\rm ordered}, ~\sum_{\alpha} \ell_{\alpha}~{\rm even},~ i=1,\cdots,\pmb{n}_{\pmb{n\ell}}\big\},
\end{eqnarray}
where $\pmb{n}_{\pmb{n\ell}}$ is the number of basis functions for the selected $(\pmb{n}, \pmb{l})$ channels; see \cite[Proposition 7 and Eq. (3.12)]{2019-ship1}. 

Once the finite symmetric polynomial basis set $\pmb{B} \subset \bigcup^{\mathcal{N}}_{N=1} \pmb{B}_N$ is constructed, the ACE site potential can be expressed as
\begin{align}\label{ships:energy}
V^{\rm ACE}(\pmb{g}; \{c_B\}_{B\in\pmb{B}}) = \sum_{B\in\pmb{B}} c_B B(\pmb{g}) 
\end{align}
with the coefficients $c_{B}$. The corresponding force of this potential is denoted by $\F^{\rm ACE}$.

The family of potentials are systematically improvable (see \cite[Section 6.2]{2019-ship1}): by increasing the body-order, cutoff radius and polynomial degree they are in principle capable of representing an arbitrary many-body potential energy surface to within arbitrary accuracy~\cite{2019-ship1, witt2023acepotentials}.


\bibliographystyle{elsarticle-num} 
\bibliography{bib.bib}





\end{document}

%% file: notation.tex
\def\per{{\rm per}}
\def\mA{{\sf A}}
\def\ulin{u^{\rm lin}}
\def\burg{{\sf b}}
\def\T{\mathcal{T}}
\def\D{\mathcal{D}}

\def\Z{\mathbb{Z}}
\def\bv{\mathsf{b}}

\def\assERL{\textnormal{\bf (RL)}}

\def\Vhom{V^{\rm h}}

\def\E{\mathcal{E}}

\def\p0{\pmb{0}}

\def\Us{\mathscr{U}}

\def\UsH{{\mathscr{U}}^{1,2}}

\def\efit{\varepsilon^{\rm E}}
\def\ffit{\varepsilon^{\rm F}}

\def\vfit{\varepsilon^{\rm V}}
\def\L{\Lambda}
\def\R{\mathbb{R}}
\def\Rl{\mathcal{R}}

\def\<{\langle}
\renewcommand{\>}{\rangle}

\def\F{\mathcal{F}}
\def\N{\mathbb{N}}

\def\dx{\,{\rm d}x}

\def\ds{\,{\rm d}s}
\def\Wcb{W_{\rm cb}}

\def\n{\mathfrak{n}}